\documentclass[preprint,12pt]{elsarticle}
\usepackage{url}
\usepackage{hyperref}
\usepackage{hyphenat}
\usepackage{microtype}
\usepackage[final]{changes}

\usepackage{amssymb}
\usepackage{amsmath}
\usepackage{booktabs}
\usepackage{pdflscape} 
\usepackage{enumitem} 
\usepackage{array} 
\usepackage{longtable} 
\usepackage[table]{xcolor} 
\usepackage{framed} 
\usepackage{tabularx}
\usepackage{ragged2e}
\usepackage{subcaption}
\usepackage{graphicx}
\usepackage{rotating}
\usepackage{tikz}
\usepackage{xurl}
\usepackage{url}
\usepackage{hyperref}
\usepackage{array}

\journal{Journal of Systems and Software}

\begin{document}

\begin{frontmatter}



\title{An Empirical Study of Generative AI Adoption in Software Engineering}


\author{Görkem Giray} 
\affiliation{organization={Eindhoven University of Technology},
            city={Eindhoven},
            country={The Netherlands}}
\affiliation{organization={Izmir Institute of Technology},
            city={Izmir},
            country={Türkiye}}

\author{Onur Demirörs} 
\affiliation{organization={Izmir Institute of Technology},
            city={Izmir},
            country={Türkiye}}

\author{Marcos Kalinowski} 
\affiliation{organization={Pontifical Catholic University of Rio de Janeiro (PUC-Rio)},
            city={Rio de Janeiro},
            country={Brazil}}

\author{Daniel Mendez} 
\affiliation{organization={Blekinge Institute of Technology},
            city={Karlskrona},
            country={Sweden}}
\affiliation{organization={fortiss},
            city={Munich},
            country={Germany}}

\begin{abstract}
\textit{Context.} Generative Artificial Intelligence (GenAI) tools are being increasingly adopted by practitioners in software engineering (SE), promising support for several SE activities. Despite increasing adoption, we still lack empirical evidence on how GenAI is used in practice, the benefits it provides, the challenges it introduces, and its broader organizational and societal implications.

\textit{Objective.} This study aims to provide an overview of the status of GenAI adoption in SE. It investigates the (1) status of GenAI adoption, (2) associated benefits and challenges, (3) institutionalization of tools and techniques, and (4) anticipated long-term impacts on SE professionals and the community.

\textit{Method.} We conducted a questionnaire-based survey distributed across 38 countries to collect insights from SE practitioners. The survey combined closed-ended and open-ended questions to capture both quantitative trends and qualitative insights. We received 204 responses from 38 countries. Qualitative responses were analyzed using systematic coding and categorization. To cope with the random sampling limitation, we used bootstrapping and conservatively reported confidence intervals.

\textit{Results.} The results indicate a wide adoption of GenAI tools and that they are deeply integrated into daily SE work, particularly for implementation, verification and validation, personal assistance, and maintenance-related tasks. Practitioners perceive substantial benefits, most notably reduction in cycle time, quality improvements, enhanced support in knowledge work, and productivity gains; however, as most organizations lack objective productivity and quality metrics, these findings reflect practitioner sentiment rather than independently validated outcomes. Significant challenges persist, including incorrect or unreliable outputs, prompt engineering difficulties, validation overhead, security and privacy concerns, and risks of overreliance. Institutionalization of tools and techniques seems to be common, but it varies considerably, with a strong focus on tool access and less emphasis on training and governance. Practitioners expect GenAI to redefine rather than replace their roles, while expressing moderate concern about possible job market contraction and skill shifts; these findings reflect current perceptions and anticipations rather than observed outcomes.

\textit{Conclusions.} The findings further corroborate that GenAI is receiving much attention in industrial SE practice, delivering perceived value, but also introducing new technical, organizational, and socio-technical challenges. Practitioners, organizations, and even governments, as we argue, must move beyond ad-hoc adoption towards more systematic approaches to ensure successful, sustainable, and responsible integration of GenAI.
\end{abstract}

\begin{graphicalabstract}
\end{graphicalabstract}

\begin{highlights}
\item \textit{Adoption is mainstream:} ~80\% use GenAI regularly, with daily or near‑daily usage common.
\item \textit{Top barriers to adoption:} ~20\% do not use GenAI mainly due to skill/time constraints, lack of perceived need and insufficient maturity of GenAI tools.
\item \textit{Top use cases:} Implementation leads (coding, code completion), followed by verification \& validation, personal assistance and maintenance.
\item \textit{Reported benefits:} Faster cycle times, stronger support for knowledge work, perceived quality improvements, and productivity gains.
\item \textit{Measurement gap:} Most teams do not use objective metrics for productivity and quality measurement. An opportunity for leaders is to standardize how impact is tracked.
\item \textit{Key challenges:} Accuracy/reliability of the outputs, prompt engineering, validation overhead, and security/privacy concerns.
\item \textit{Organizational readiness:} Many organizations provide access to tools; fewer invest in training, clear policies, or defined objectives/KPIs. Moving beyond ad‑hoc adoption is essential for organizations.
\end{highlights}


\end{frontmatter}



\section{Introduction} \label{introduction}
Generative Artificial Intelligence (GenAI) refers to the subfield of AI that can create new content, such as text, images, video, audio, or code. GenAI relies on deep learning models typically trained on a very large scale of data. The capabilities of GenAI tools for creating new content across a broad range of use cases often comes with the promise to transform several industries, including Software Engineering (SE). The adoption of GenAI applications seems to be growing rapidly, but unevenly across industries, and the information technology sector is meant to be pioneering \cite{WEF2025, OpenAI2025}. As OpenAI reports, users with engineering roles are using ChatGPT mainly for programming tasks, followed by research and documentation \cite{OpenAI2025}. A recent more rigorous analysis of four million Claude interactions revealed that it is predominantly used in software development, mostly by developers, testers, and analysts \cite{Handa2025}.

Further industrial research initiatives, such as the one by the Capgemini Research Institute \cite{Capgemini2024}, measured GenAI adoption among SE professionals at 46\% in 2024, forecasting it to reach 85\% in 2026. The StackOverflow Developer Survey \cite{StackOverflow2025} reports that 84\% of its respondents are already using or planning to use AI tools, and 51\% of professional developers reported daily use. SE professionals use GenAI tools mainly for coding and related activities \cite{McKinsey2023, Capgemini2024, OpenAI2025}. The use of GenAI tools as an advanced search engine is also one of the main use cases \cite{StackOverflow2025}. The existing and potential use cases are much wider across the software development life cycle \cite{Ozkaya2023, Pezze2025}.

There are also significant challenges associated with the adoption of GenAI tools in software engineering, which in some cases lead users to refrain from using these tools altogether. GenAI tools may produce code that does not meet functional requirements or quality expectations \cite{Vaithilingam2022, Liang2024} and includes security vulnerabilities \cite{Khemka2024, Martinovic2025}. Moreover, considerable time may be required to debug and modify imperfect GenAI-generated code \cite{Vaithilingam2022, Liang2024}. Furthermore, developers worry about privacy risks related to sharing proprietary code with public models and the potential for the generated code to infringe on intellectual property \cite{Wang2024, Liang2024}. Finally, GenAI may have long-term impacts on SE professionals and the community, e.g., skills atrophy \cite{Wang2024}, job displacements \cite{StackOverflow2025}, and a potential contraction of the SE job market \cite{Challapally2025}.

The complex and intertwined technical, human, and organizational factors impacting the adoption of GenAI for SE are reflected in the so far scattered literature. The lack of evidence-based insights into the extent to which GenAI is adopted and the tasks in scope of GenAI, as well as the challenges associated with GenAI, renders problem-driven research in this area cumbersome. This motivated our survey initiative at hands, establishing a baseline for the effective utilization and implementation of GenAI tools by users, organizations, and societies. Our study uses empirical data collected from 204 SE practitioners to address these critical themes and provide a holistic view of the dynamic landscape of GenAI adoption in SE. The remainder of this paper is organized as follows. Section \ref{backgroundandrelatedwork} provides background and related work. Section \ref{researchmethod} describes the research method. Section \ref{results} presents the results of the survey. Sections \ref{discussion} and \ref{threatstovalidity} discuss the results and threats to validity, respectively. Finally, section \ref{conclusions} concludes the paper.

\section{Background and Related Work} \label{backgroundandrelatedwork}
This section introduces the foundational concepts and related terms as used in this paper. Section \ref{relatedwork} presents the related work from the academic and gray literature.

\subsection{Background} \label{background}
This section introduces the foundational concepts and describes the vocabulary by defining the essential concepts used in this paper. Section \ref{relatedwork} presents the related work from the academic and gray literature.

\subsubsection{AI and GenAI} \label{aiandgenai}
Artificial Intelligence (AI) is the discipline devoted to the development of intelligent agents \cite{Russell2013}. Intelligent agents perceive their environment and try to achieve their goals by acting autonomously \cite{Russell2013}. Generative AI or GenAI refers to the subfield of AI that can create new content, such as text, images, video, audio, or code. GenAI relies on deep learning models trained on a very large scale of data.

GenAI gained significant attention after the introduction of ChatGPT by OpenAI in November 2022. Other chatbots, such as Google's Gemini, Anthropic’s Claude, DeepSeek Chat, and Perplexity, launched soon after and reached a significant user base. GenAI tools specialized in SE-related tasks were also launched to support SE professionals. For example, GitHub Copilot and Cursor provide GenAI capabilities, such as code generation, intelligent code completion, codebase understanding, and smart rewrites, to SE practitioners. Additionally, pre-trained language models can be used for specific SE tasks, such as software size measurement \cite{Unlu2026, Tenekeci2026}.

\subsubsection{SE for GenAI and GenAI for SE} \label{seforgenaiandgenaiforse}
The interplay between the SE and AI disciplines dates back decades. Rech and Althoff \cite{Rech2004} present examples of methods and techniques borrowed from one discipline to support practice and research in the other. Partridge \cite{Partridge1988} underlines the importance of the development environment to develop AI-enabled systems and shared ideas on the application of AI to SE, i.e., automatic programming. As seen in these examples, two areas of synergy bring together the knowledge and experience of these two disciplines \cite{Giray2021a}:

\begin{itemize}
    \item \textit{SE for GenAI} refers to the addressing of various SE tasks for engineering GenAI systems or hybrid systems involving GenAI components. Here, researchers aim at identifying and adopting various aspects relevant to the engineering of (Gen)AI-enabled systems compared to traditional deterministic software. Hassan et al. \cite{Hassan2024a}, for instance, discuss the challenges of developing software that uses Foundation Models (FMs), such as Large Language Models (LLMs), as one of its components (such software is named FMware). They identified some revisions to traditional SE practices to develop FMware, such as cognitive architecture planning and design. Meanwhile, there exist various approaches towards SE for (Gen)AI and although isolated, they already provide valuable grounds for adoption. Kästner provides a reading list \cite{Kästner2024} used in the context of higher education to which we refer in this context.
    \item \textit{GenAI for SE} refers to the application or adaptation of GenAI tools to address various SE tasks \cite{Hou2024}, such as drafting software requirements specifications \cite{Krishna2024}, code generation \cite{Jiang2024}, code comment generation \cite{Geng2024}, and test case generation \cite{Zhang2023}. Researchers aim to utilize GenAI capabilities to engineer software more efficiently and effectively.
\end{itemize}

This study focuses on GenAI for SE activities.

\subsection{Related Work} \label{relatedwork}
The topic is gaining much attention from the research community with several studies carried out to understand the impact of GenAI tools on SE. Table~\ref{tab:selected_existing_studies} provides an overview of existing questionnaire-based studies on the adoption of GenAI in SE.

Industry reports indicate that 10-16\% of practitioners do not use GenAI tools for SE tasks \cite{Google2025b, DefineX2025, StackOverflow2025}. However, these reports do not provide specific data, percentages, or categorical reasons explaining why non-users explicitly choose to abstain from GenAI tools. Only two papers, i.e., \cite{Liang2024} and \cite{Sergeyuk2025}, provide motivations for not using AI programming assistants. Our study provides a list of categorized reasons for not using GenAI tools for SE tasks.

From an SE coverage perspective, most previous work focuses on all knowledge areas \cite{Banh2025, DeCampos2024, Glushkova2023, Khemka2024, Russo2024a, Ulfsnes2024, Capgemini2024, DefineX2025} or implementation \cite{Coutinho2024, Haque2024, Hassan2024b, Kuhail2024, Li2024, Liang2024, Sergeyuk2025, Ziegler2022}. \added{Afroz et al. surveyed developers to understand the perceived impact of GenAI tools on developer productivity \cite{afroz2026fast}.} Implementation-related activities, e.g., code generation, code translation, and prototyping, consistently emerge as the dominant use case \cite{Google2025b, Liang2024, Kuhail2024, Banh2025, DeCampos2024, felder2026adoption}, a pattern corroborated by OpenAI's analysis of ChatGPT usage among engineering roles \cite{OpenAI2025}. Maintenance activities such as refactoring and legacy modernization \cite{Google2025b, Capgemini2024}, verification and validation, including test case generation \cite{Khemka2024, Sergeyuk2025, Martinovic2025}, and requirements engineering support \cite{Capgemini2024, DefineX2025, Glushkova2023, Google2025b} are also reported across multiple sources. Our study covers all SE knowledge areas using a standardized ISO/IEC/IEEE 12207 taxonomy that enables a lifecycle-wide empirical picture.

Industrial reports consistently project large productivity gains from GenAI adoption. The Capgemini Research Institute \cite{Capgemini2024} estimates organizational productivity improvements of 7–18\% across the software development lifecycle, and reports that AI can reduce time spent resolving coding issues by up to 30\%. The DefineX, a consulting house and venture builder, survey reports approximately 10\% average time savings \cite{DefineX2025}, while data from DX, an organization that provides a developer intelligence platform, suggest that developers save an average of 3.6 hours per week \cite{Laura2025}. However, academic evidence presents a more nuanced and sometimes contradictory picture. Peng et al. \cite{Peng2023} found that GitHub Copilot enabled developers to complete a standardized programming task 55.8\% faster; it is also important to note that this result was obtained in a controlled setting with a specific task type. Song et al. \cite{Song2024} found an improvement in productivity at the project level signaled by increased merged pull requests and improved individual productivity, but did not assess code quality. Critically, Becker et al. \cite{Becker2025} found that AI systems can slow down experienced developers by shifting effort from active coding to prompting, waiting, and reviewing outputs; a finding that directly contradicts the dominant narrative and suggests that productivity gains may be task-specific, experience-dependent, and context-sensitive. Most productivity claims are based on self-reported perceptions rather than objective measurement. Our study empirically documents this limitation and explicitly discusses it.

The operational gains are counterbalanced in the literature by challenges that scale along with the complexity of the task \cite{Banh2025, Google2025b, Hassan2024b, Kuhail2024, Martinovic2025}. While GenAI tools drastically reduce initial development time, their non-deterministic nature introduces high error rates, with empirical assessments showing incorrect code generation varying from 31\% to 65\% depending on model architecture \cite{Martinovic2025}. This error rate is amplified by task complexity; as problem specifications expand, the failure probability of model output scales proportionally \cite{ Kuhail2024}. GenAI tools struggle in establishing a project-specific context, such as internal APIs, business logic, architectural history, and team conventions \cite{Banh2025, Hassan2024b, Jahic2024, Sergeyuk2025}. Furthermore, empirical code reviews demonstrate that AI-assisted developers are significantly more prone to introducing exploitable vulnerabilities (up to 67\% probability) than unassisted developers (27\%) \cite{Martinovic2025}. Our study maps practitioner impediments directly to the formal characteristics of the ISO/IEC 25059:2023 standard (Quality model for AI systems). This standard-driven mapping elevates the discourse from an ad-hoc inventory of user complaints to a reproducible, theory-grounded taxonomy of systemic vulnerabilities.

\newcommand{\nocover}{\(\circ\)}
\newcommand{\cover}{\(\bullet\)}
\newcommand{\partcover}{\(\circleddash\)}
\newcommand*\emptycirc[1][0.7ex]{\tikz\draw (0,0) circle (#1);} 
\newcommand*\halfcirc[1][0.7ex]{%
  \begin{tikzpicture}
  \draw[fill] (0,0)-- (90:#1) arc (90:270:#1) -- cycle ;
  \draw (0,0) circle (#1);
  \end{tikzpicture}}
\newcommand*\fullcirc[1][0.7ex]{\tikz\fill (0,0) circle (#1);}

\newcolumntype{G}{>{\raggedright\arraybackslash}p{6.2cm}}

\begin{landscape}

\begin{table*}[t]
\caption{A list of selected studies involving findings drawn from a questionnaire \begin{scriptsize}(\textit{Data Period:} data collection period; \textit{Geographical Coverage:} countries in which the respondents are based; \textit{\# of Resp.:} number of respondents; \textit{SE \& GenAI Cover.:} Coverage of SE knowledge areas and GenAI tools investigated; \textit{Non-use Reasons:} reasons for not using GenAI tools; \textit{Use cases:} SE tasks GenAI tools are used for; \textit{GenAI Tools:} GenAI tools used for SE tasks; \textit{Usage Patterns:} reported usage patterns of GenAI tools; \textit{Benefits:} reported benefits of GenAI usage; \textit{Productivity:} perceived productivity change caused by GenAI usage; \textit{Quality:} perceived quality change caused by GenAI usage; \textit{Measurement:} metrics used for size, productivity, and quality measurement; \textit{Challenges:} reported challenges of GenAI usage; \textit{Institutionalization:} organizational governance and support; \textit{Impacts:} potential GenAI impact on SE roles, job market, skills, compensation, and social interaction; \textit{Not reported:} Data collection period or geographical coverage is not reported by the corresponding study.) \end{scriptsize}}
\label{tab:selected_existing_studies}
\centering
\footnotesize
\setlength{\tabcolsep}{3pt}
\renewcommand{\arraystretch}{1.15}

\begin{scriptsize}
\begin{tabular}{lcGccccccccccccc}
\toprule
\textbf{Study} &
\textbf{Data Period} &
\textbf{Geographical Coverage} &
\begin{sideways}\textbf{\# of Resp.}\end{sideways} &
\textbf{SE \& GenAI Cover.} &
\begin{sideways}\textbf{Non-use Reasons}\end{sideways} &
\begin{sideways}\textbf{Use Cases}\end{sideways} &
\begin{sideways}\textbf{GenAI Tools}\end{sideways} &
\begin{sideways}\textbf{Usage Patterns}\end{sideways} &
\begin{sideways}\textbf{Benefits}\end{sideways} &
\begin{sideways}\textbf{Productivity}\end{sideways} &
\begin{sideways}\textbf{Quality}\end{sideways} &
\begin{sideways}\textbf{Measurement}\end{sideways} &
\begin{sideways}\textbf{Challenges}\end{sideways} &
\begin{sideways}\textbf{Institutionalization}\end{sideways} &
\begin{sideways}\textbf{Impacts}\end{sideways} \\
\midrule
\cite{afroz2026fast} & Not reported & Not reported & 415 & SE/Impl. \& GenAI & \emptycirc & \halfcirc & \emptycirc & \fullcirc & \emptycirc & \fullcirc & \emptycirc & \halfcirc & \fullcirc & \emptycirc & \emptycirc \\
\cite{Banh2025} & Aug’23--Jan’24 & Europe (17 companies) & 18   & SE \& GenAI & \emptycirc & \emptycirc & \emptycirc & \emptycirc & \fullcirc & \emptycirc & \emptycirc & \emptycirc & \fullcirc & \emptycirc & \emptycirc \\
\cite{Coutinho2024} & Last months of 2023 & Not reported & 13   & Impl. \& GenAI & \emptycirc & \fullcirc & \emptycirc & \emptycirc & \fullcirc & \fullcirc & \emptycirc & \emptycirc & \fullcirc & \emptycirc & \emptycirc \\
\cite{Damyanov2024} & Apr’24 & Bulgaria & 104  & SE/Impl. \& GenAI & \emptycirc & \emptycirc & \emptycirc & \fullcirc & \fullcirc & \emptycirc & \emptycirc & \emptycirc & \emptycirc & \emptycirc & \halfcirc \\
\cite{DeCampos2024} & Not reported & Brazil & 45   & SE \& GenAI & \emptycirc & \fullcirc & \fullcirc & \fullcirc & \fullcirc & \emptycirc & \emptycirc & \emptycirc & \fullcirc & \emptycirc & \emptycirc \\
\cite{felder2026adoption} & Apr'25--Aug'25 & Germany & 109 & Seven SE tasks \& GenAI & \emptycirc & \halfcirc & \fullcirc & \fullcirc & \fullcirc & \fullcirc & \emptycirc & \emptycirc & \fullcirc & \emptycirc & \emptycirc \\
\cite{Glushkova2023} & Mar’23--Jul’23 & Russia, Italy, Germany, USA, etc. & 150  & SE \& ChatGPT & \emptycirc & \emptycirc & \emptycirc & \fullcirc & \fullcirc & \fullcirc & \fullcirc & \emptycirc & \emptycirc & \emptycirc & \emptycirc \\
\cite{Haque2024} & Not reported & Not reported & 128  & Impl. \& AI Assist. & \emptycirc & \fullcirc & \emptycirc & \fullcirc & \fullcirc & \fullcirc & \halfcirc & \emptycirc & \emptycirc & \emptycirc & \halfcirc \\
\cite{Hassan2024b} & Not reported & Bangladesh, Singapore, UK, Germany, Finland, Estonia, Japan & 11   & Impl. \& AI Tools & \emptycirc & \emptycirc & \emptycirc & \fullcirc & \fullcirc & \emptycirc & \emptycirc & \emptycirc & \emptycirc & \emptycirc & \emptycirc \\
\cite{Jahic2024} & Not reported & Austria, Bosnia and Herzegovina, Germany, Serbia, UK & 15   & SE/Arch. \& LLMs & \emptycirc & \emptycirc & \emptycirc & \fullcirc & \emptycirc & \emptycirc & \emptycirc & \emptycirc & \fullcirc & \emptycirc & \emptycirc \\
\cite{Khemka2024} & Apr’23 & USA & 737  & SE \& AI Tools & \emptycirc & \emptycirc & \emptycirc & \fullcirc & \emptycirc & \emptycirc & \emptycirc & \emptycirc & \fullcirc & \emptycirc & \emptycirc \\
\cite{Kuhail2024} & Not reported & UK, USA, UAE, etc. & 99   & Impl. \& ChatGPT & \emptycirc & \emptycirc & \emptycirc & \fullcirc & \fullcirc & \emptycirc & \emptycirc & \emptycirc & \fullcirc & \emptycirc & \emptycirc \\
\cite{Li2024} & Summer 2023 & Not reported & 15   & Impl. \& GenAI & \emptycirc & \emptycirc & \emptycirc & \emptycirc & \fullcirc & \emptycirc & \emptycirc & \emptycirc & \emptycirc & \emptycirc & \emptycirc \\
\cite{Liang2024} & Jan’23 & 57 countries & 410  & Impl. \& Prog. Assist. & \fullcirc & \fullcirc & \fullcirc & \fullcirc & \fullcirc & \emptycirc & \emptycirc & \emptycirc & \fullcirc & \emptycirc & \emptycirc \\
\cite{Martinovic2025} & Feb’24 & Bosnia \& Herzegovina, Croatia, etc. & 84   & Quality \& AI Tools & \emptycirc & \emptycirc & \fullcirc & \fullcirc & \fullcirc & \emptycirc & \emptycirc & \emptycirc & \emptycirc & \emptycirc & \halfcirc \\
\cite{Russo2024a} & 2023 & Portugal, South Africa, Italy, UK, Poland, etc. & 100  & SE \& GenAI & \emptycirc & \fullcirc & \emptycirc & \emptycirc & \fullcirc & \fullcirc & \emptycirc & \emptycirc & \fullcirc & \emptycirc & \halfcirc \\
\cite{Sergeyuk2025} & Not reported & 71 countries & 481  & Impl. \& Prog. Assist. & \fullcirc & \fullcirc & \fullcirc & \fullcirc & \emptycirc & \emptycirc & \emptycirc & \emptycirc & \emptycirc & \emptycirc & \emptycirc \\
\cite{Ulfsnes2024} & Not reported & Not reported & 13   & SE \& GenAI & \emptycirc & \fullcirc & \emptycirc & \emptycirc & \fullcirc & \emptycirc & \emptycirc & \emptycirc & \fullcirc & \emptycirc & \emptycirc \\
\cite{Ziegler2022} & Feb’22--Mar’22 & Not reported & 2047 & Impl. \& Copilot & \emptycirc & \emptycirc & \emptycirc & \emptycirc & \fullcirc & \emptycirc & \halfcirc & \emptycirc & \emptycirc & \emptycirc & \emptycirc \\
\midrule
\textbf{This} & May'25--Nov'25 & 38 countries & 204  & SE \& GenAI & \fullcirc & \fullcirc & \fullcirc & \fullcirc & \fullcirc & \fullcirc & \fullcirc & \fullcirc & \fullcirc & \fullcirc & \fullcirc \\
\bottomrule
\end{tabular}

\vspace{2pt}
\footnotesize{\emptycirc\ Not covered;\quad \fullcirc\ Covered;\quad \halfcirc\ Partially covered.}
\end{scriptsize}
\end{table*}

\end{landscape}

\section{Research Method} \label{researchmethod}
This section describes our research design by presenting the research questions in Section \ref{goalandresearchquestions}, the questionnaire development and its overall structure in Sections \ref{questionnairedevelopment} and \ref{questionnairestructure}, the detailed explanation of the data collection and analysis procedures in Sections \ref{datacollection} and \ref{dataanalysis}.

\subsection{Goal and Research Questions} \label{goalandresearchquestions}
The purpose of this study is to understand the adoption of GenAI tools in SE activities from the point of view of SE professionals. We aim at characterizing the context (i.e., country, experience, education, role, sector, size of organization and team, and project management approach), the status of adoption, and the benefits and challenges of using GenAI tools in SE. Based on this goal, we raise the following research questions (RQs):
\bigskip

RQ1. What is the status of GenAI tool use in SE?

This question addresses the SE tasks GenAI tools are used for, the GenAI tools utilized, their usage patterns, and institutionalization. In addition, this question seeks to determine the reasons behind not using GenAI tools. This RQ is refined into more detailed questions as follows:

\bigskip
\begin{itemize}
\item RQ1.1. What are the reasons behind not using GenAI tools?
\item RQ1.2. What are the SE tasks for which SE professionals use GenAI tools?
\item RQ1.3. Which GenAI tools are used and what are the usage patterns?
\end{itemize}

\bigskip

RQ2. What are the benefits and challenges of using GenAI tools for SE tasks?

This question aims to understand the reported benefits of GenAI tools, perceived productivity change, measurement approaches used, as well as the barriers to obtaining benefits. This RQ is refined into more detailed questions as follows:

\bigskip
\begin{itemize}
\item RQ2.1. What are the reported benefits?
\item RQ2.2. What is the perceived productivity change?
\item RQ2.3. What is the perceived change in the quality of SE artifacts?
\item RQ2.4. How are size, productivity, and quality measured?
\item RQ2.5. What are the challenges of using GenAI tools for SE tasks?
\end{itemize}

\bigskip

RQ3. What is the status of institutionalization for GenAI tool use in SE?

This question aims to understand how organizations are governing GenAI adoption.
\bigskip

RQ4. What are the expected impacts of GenAI tools on the SE community?

This question aims to understand the social and professional impact of GenAI on the SE community.

\subsection{Questionnaire Development} \label{questionnairedevelopment}
We designed our survey based on best practices of survey research \cite{Molleri2020, Wagner2020} and previous experience \cite{Kalinowski2025, Garousi2019a, Garousi2015, Giray2021b, Akdur2018}. The questionnaire was initially developed by the first two authors and subsequently refined through an iterative pilot study. Five experienced SE professionals completed the preliminary version, providing crucial feedback on clarity, structure, and scope. Following revisions based on this input, the final instrument was validated by a social scientist specializing in survey methodology. The Research Ethics Committee at Izmir Institute of Technology approved the questionnaire. The questionnaire was then implemented on the surveyjs.io platform.

\subsection{Questionnaire Structure} \label{questionnairestructure}
The questionnaire consists of six sections as listed in Table~\ref{tab:list_of_survey_questions}. The questionnaire starts with an introduction and screening question. The introduction includes the main objective, a short definition, and examples of GenAI tools and SE activities. Participants started with the questionnaire if they answered the screening question (C1) as “yes”, otherwise they were directed to the “thank you” page without answering any question. The Demographics section includes nine questions to describe the sample. The next section includes ten questions to understand the status of GenAI use in SE as well as associated benefits and challenges. The fourth section includes two questions to get the picture of institutionalization of tools and techniques. The fifth section encompasses seven Likert scale items to provide a foresight for the potential impacts of GenAI tools on SE community. The questionnaire concludes with two optional questions asking for additional critical issues participants would like to share and email address.

\begin{landscape}
\scriptsize
\begin{longtable}{
    >{\raggedright\arraybackslash}p{2.2cm} 
    >{\raggedright\arraybackslash}p{0.7cm} 
    >{\raggedright\arraybackslash}p{9.0cm} 
    >{\raggedright\arraybackslash}p{2.0cm} 
    >{\raggedright\arraybackslash}p{2.0cm} 
    >{\raggedright\arraybackslash}p{0.7cm} 
}
\caption{List of survey questions \label{tab:list_of_survey_questions}} \\
\toprule
\textbf{Section} & \textbf{No} & \textbf{Question} & \textbf{Type} & \textbf{Condition} & \textbf{RQ} \\ \midrule
\endfirsthead

\multicolumn{6}{c}%
{{\bfseries \tablename\ \thetable{} -- continued from previous page}} \\
\toprule
\textbf{Section} & \textbf{No} & \textbf{Question} & \textbf{Type} & \textbf{Condition} & \textbf{RQ} \\ \midrule
\endhead

\midrule
\multicolumn{6}{r}{{Continued on next page}} \\
\bottomrule
\endfoot

\bottomrule
\endlastfoot

Consent & C1 & Do you agree to participate in the survey?  & Yes/No & & \\ \midrule
Demographics & D2 & In which country do you live?  & Closed (Single) & If C1 = Yes & \\
 & D3 & Please select your highest formal education degree relevant to your role.  & Closed (Single) & & \\
 & D4 & Please select your highest formal education field of study relevant to your role.  & Closed (Single) & & \\
 & D5 & How many years of professional experience do you have in software engineering?  & Number & & \\
 & D6 & Which roles/positions best describe your current activities within your organization?  & Closed (Multiple) & & \\
 & D7 & Please select the main industrial sector for which you engineer software.  & Closed (Single) & & \\
 & D8 & What is the size of the organization you currently work for?  & Closed (Single) & & \\
 & D9 & What is the size of your team?  & Closed (Single) & & \\
 & D10 & How would you describe the project management approach in your organization?  & 5-point Likert & & \\ \midrule
GenAI Tool Use & Q11 & Are you using GenAI tools for your SE activities?  & Yes/No & & 1 \\
 & Q12 & What is(are) the reason(s) of not using GenAI tools?  & Free text & If Q11 = No & 1.1 \\
 & Q13 & Which specific SE activity are you using GenAI tool(s) for?  & Free text & If Q11 = Yes & 1.2 \\
 & Q14 & Which GenAI tool(s) do you use for \{SE ACTIVITY\}?  & Free text & & 1.3 \\
 & Q15 & How frequently do you use GenAI tools for \{SE ACTIVITY\}?  & 5-point Likert & & 1.3 \\
 & Q16 & What benefit(s) do you encounter by using GenAI tools for \{SE ACTIVITY\}?  & Free text & & 2.1 \\
 & Q17 & Approximately how much time do you spend to finish an activity that used to take 8 hours without GenAI tools?  & 5-point Likert & & 2.2 \\
 & Q18 & GenAI tools enable me to achieve better quality of my work for \{SE ACTIVITY\}.  & 5-point Likert & & 2.3 \\
 & Q19 & Do you measure size, productivity and quality using an objective metric?  & Free text & & 2.4 \\
 & Q20 & What challenge(s) do you face when using GenAI tools for \{SE ACTIVITY\}?  & Free text & & 2.5 \\ \midrule
Institutional-ization & Q21 & Does your organization provide any support to use GenAI tool(s) for SE activities?  & Yes/No & & 3 \\
 & Q22 & What is your organization's approach to GenAI tool use for SE activities?  & Closed (Multiple) & If Q21 = Yes & 3 \\ \midrule
Potential Impact & Q23 & GenAI will replace my current role within the next five years.  & 5-point Likert & & 4 \\
 & Q24 & GenAI will redefine my current role rather than replace within the next five years.  & 5-point Likert & & 4 \\
 & Q25 & GenAI will decrease the number of available SE-related jobs within the next five years.  & 5-point Likert & & 4 \\
 & Q26 & GenAI will make the majority of my current skills obsolete within the next five years.  & 5-point Likert & & 4 \\
 & Q27 & I am confident in my ability to acquire the necessary skills to effectively integrate GenAI.  & 5-point Likert & & 4 \\
 & Q28 & GenAI will result in decreased compensation and benefits within the next five years.  & 5-point Likert & & 4 \\
 & Q29 & GenAI will result in less social interactions in the workplace within the next five years.  & 5-point Likert & & 4 \\ \midrule
Additional Information & Q30 & Please provide any other important issues about GenAI use in SE.  & Free text & & \\
 & Q31 & Provide your email address to receive an overview of results (optional).  & Free text & & \\
\end{longtable}
\end{landscape}

\subsection{Data Collection} \label{datacollection}

The target population concerns professionals performing SE-related activities, such as requirements engineers, architects, developers, testers, project managers, etc. We employed three non-probability techniques:

\begin{itemize}
    \item \textit{Convenience sampling}, which refers to selecting participants based on availability or expedience \cite{Baltes2022}. Convenience sampling is used widely in SE research mostly due to speed and low cost, including large-scale surveys such as \cite{Ralph2020} and \cite{Kalinowski2025}. We distributed the survey to our professional network through emails and LinkedIn.
    
    \item \textit{Purposive sampling}, which refers to selecting some logic or strategy, carefully but not randomly \cite{Baltes2022}. In addition to convenience sampling, we identified potential participants from our personal network residing in different countries. Moreover, we have intentionally sent the questionnaire to at most two participants from the same organization. The rationale behind this purposive sampling is to have a heterogeneous sampling across countries and organizations, which may improve generalization from a cultural and organizational perspective.
    
    \item \textit{Snowball sampling}, which refers to selecting participants based on their relationship to previously selected participants \cite{Baltes2022}, i.e., to the participants we selected through convenience and purposive sampling in our case. Snowballing allowed us to reach more people in different countries through our contacts.
\end{itemize}

The units of analysis in this survey are software professionals working in the industry who individually and anonymously participated in the online survey. Therefore, all results reported in this paper are tied to the software professionals under investigation, not to organizations or projects.

The questionnaire was available to participants from May to November 2025. We received 223 responses, of which 204 were used for analysis.

\subsection{Data Analysis} \label{dataanalysis}
We conducted a quality check on the responses before starting with data analysis. We removed ten responses explicitly declining consent to complete the questionnaire, i.e., question C1 responded as “no”. For the responses obtained from participants who declared GenAI tool use for SE, i.e., Q11 is responded as “yes”, we checked whether a valid SE activity for which GenAI tools are used is provided. We removed nine responses that did not meet this criterion. Lastly, we checked the completeness of the responses and verified that all remaining responses are complete.

We used inferential statistics to analyze the responses provided for the third, fourth, and fifth sections since the population, i.e., software professionals, has an unknown theoretical distribution. In such cases, resampling approaches, such as bootstrapping, are considered more reliable than drawing inferences directly from the sample \cite{Wagner2020, Lunneborg2001}. Therefore, we used bootstrapping to compute confidence intervals for our findings, following the approach used in other large-scale survey studies, such as that of Wagner et al. \cite{Wagner2019} and Kalinowski et al. \cite{Kalinowski2025}. It is important to note that the confidence intervals represent within-sample variability, not population-level variability, since we bootstrap with resampling from the same non-probability sample.

Briefly, bootstrapping consists of repeatedly resampling – with replacement – from the original dataset and computing the statistic of interest for each resample. For each question, we take the N valid responses and generate S bootstrap resamples (with replacement), each of size N. Consistent with the recommendations of the statistics literature, N corresponds to the number of valid responses for the question \cite{Efron1994}, and we set S = 1,000, a commonly used value that provides stable estimates \cite{Lei2003}.

We conducted a qualitative analysis using coding procedures from grounded theory \cite{Stol2016} for open-text questions, i.e., Q12, Q13, Q14, Q16, Q19, and Q20. First, the original responses are split into atomic parts so that each part is labeled with one code.

For question Q13, i.e., SE activities GenAI tools used for, we used axial coding and mapped the atomic activities (such as coding, testing, troubleshooting) to a standardized taxonomy based on ISO/IEC/IEEE 12207 (Systems and software engineering – Software life cycle processes) \cite{ISO12207}. While the standard provides a high-level framework, we adapted our coding approach to reflect the variations and level of detail provided by the respondents.

\begin{itemize}
    \item We aggregated two categories when the reported level of detail is not sufficient to distinguish them: (1) “Stakeholder Needs and Requirements Definition” and “System/Software Requirements Definition” were merged into “Requirements Definition”, (2) “Verification” and “Validation” were consolidated as one category and named “Verification \& Validation”, (3) “Acquisition” and “Supply” were merged into a single category to categorize one response.

    \item We considered additional context provided with general-purpose activities, such as documentation or knowledge acquisition. For instance, if a respondent specified the subject of the documentation (e.g., “documenting code”, “writing comments,” or “specification document preparation”), the activity was mapped to the corresponding technical process (e.g., Implementation or Requirements Definition). Conversely, generic mentions of documentation were classified under the technical management processes of “Information Management”. Similarly, the activities related to learning were coded based on their application. For example, “programming tutorial” or “help with software development questions” were classified under Implementation, as learning activities facilitate coding.

    \item A considerable portion of the activities involved cognitive offloading and support that do not map to a specific process in the standard. To capture such activities, we established a new category titled Personal Assistance to label activities such as brainstorming, knowledge search, learning, and problem solving.
\end{itemize}

For the rest of the questions, we conducted open coding in cycles. In the initial cycles, we identified emerging patterns of similarity or contradiction. Afterwards, we collapsed and expanded the codes to understand any patterns. After we extracted the main themes and codes, we revised the codes assigned to each atomic item. The first author did the initial cycles. Then, the main themes and codes were established via discussions between the first and second authors. Additionally, for question Q20, i.e., challenges of using GenAI tools in SE, the first author mapped the challenge codes to quality characteristics in the ISO/IEC 25059:2023 standard (Quality model for AI systems) \cite{ISO25059}. The first and second authors then discussed and finalized these mappings for validation.

All codes were validated by the third and fourth authors. During validation, mistakes were corrected in splitting the responses into atomic parts and some codes were revised. For instance, a code labeling concerns on security, privacy, and copyright was split into three codes, i.e., concerns on security, privacy and ethics, and copyright and IP, for better clarity. All authors resolved disagreements through discussions and reached a consensus. We did not compute any interreviewer reliability metric since we sought consensus for all codes, as applied in other studies \cite{kafoe2025making}.

\ref{sec:appendix} provides three tables for selected participant responses and the categories assigned to SE tasks, benefits, and challenges.

\section{Results} \label{results}
In the following, we describe the sample demographics before reporting on the results according to the four research questions as introduced in Section \ref{goalandresearchquestions}.

\subsection{Sample Demographics}
\begin{figure}[h]
  \centering
  \includegraphics[width=\linewidth]{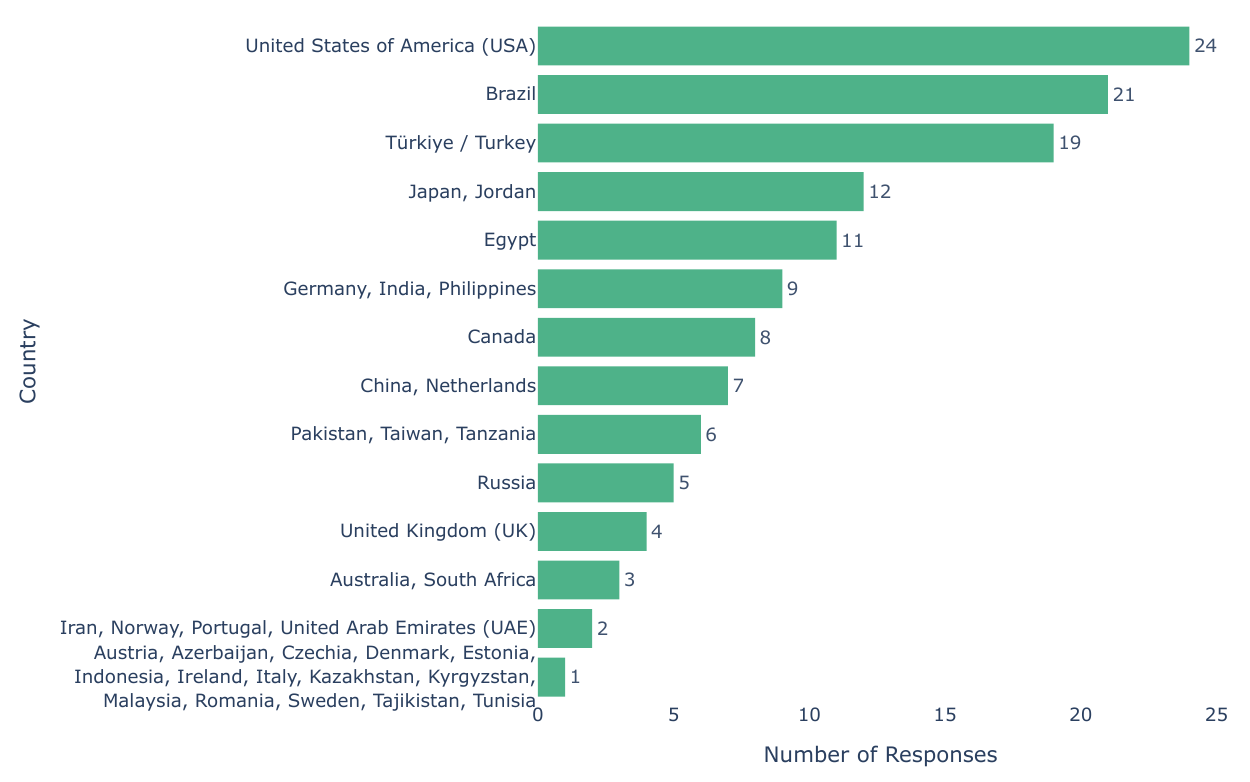}
  \caption{Number of responses per country (\textit{n} = 204)}
  \label{figure01}
\end{figure}

\textit{Country.} As Figure \ref{figure01} shows, we have respondents from 38 unique countries. The largest proportion of respondents originates from the USA (24), Brazil (21), and Türkiye (19). Other major contributors include Japan (12), Jordan (12), and Egypt (11). Five or more participants responded to the questionnaire from ten countries, i.e., Germany, India, Philippines, Canada, China, Netherlands, Pakistan, Taiwan, Tanzania, and Russia.

Table \ref{tab:responses_by_regions} provides a regional breakdown of respondents using World Bank regions (https://data.worldbank.org/country). While the sample spans seven broad world regions, the distribution is uneven: Europe \& Central Asia, East Asia \& Pacific, Middle East \& North Africa, and North America are more strongly represented. Latin America \& Caribbean region is represented by practitioners only from Brazil. South Asia and Sub-Saharan Africa are represented by only a small number of respondents. This distribution should be interpreted as evidence of geographic diversity rather than regional statistical representativeness.

\begin{table}[ht]
\caption{Distribution of responses by regions defined by World Bank (\textit{n} = 204)}
\label{tab:responses_by_regions}
\scriptsize
\begin{tabular}{p{0.27\textwidth} p{0.50\textwidth} r r}
\toprule
\textbf{Region} & \textbf{Countries} & \textbf{\textit{n}} & \textbf{\%} \\
\midrule
East Asia \& Pacific
  & Australia, China, Indonesia, Japan, Malaysia, Philippines, Taiwan
  & 39 & 19.1 \\
\addlinespace[3pt]
Europe \& Central Asia
  & Austria, Azerbaijan, Czechia, Denmark, Estonia, Germany, Ireland,
    Italy, Kazakhstan, Kyrgyzstan, Netherlands, Norway, Portugal,
    Romania, Russia, Sweden, Tajikistan, Türkiye, United Kingdom
  & 60 & 29.4 \\
\addlinespace[3pt]
Latin America \& Caribbean
  & Brazil
  & 21 & 10.3 \\
\addlinespace[3pt]
Middle East \& North Africa
  & Egypt, Iran, Jordan, Pakistan, Tunisia, United Arab Emirates
  & 34 & 16.7 \\
\addlinespace[3pt]
North America
  & Canada, United States of America
  & 32 & 15.7 \\
\addlinespace[3pt]
South Asia
  & India
  & 9 & 4.4 \\
\addlinespace[3pt]
Sub-Saharan Africa
  & South Africa, Tanzania
  & 9 & 4.4 \\
\midrule
\textbf{Total} & & \textbf{204} & \textbf{100.0} \\
\bottomrule
\end{tabular}
\end{table}

\begin{figure}[h]
  \centering
  \includegraphics[width=\linewidth]{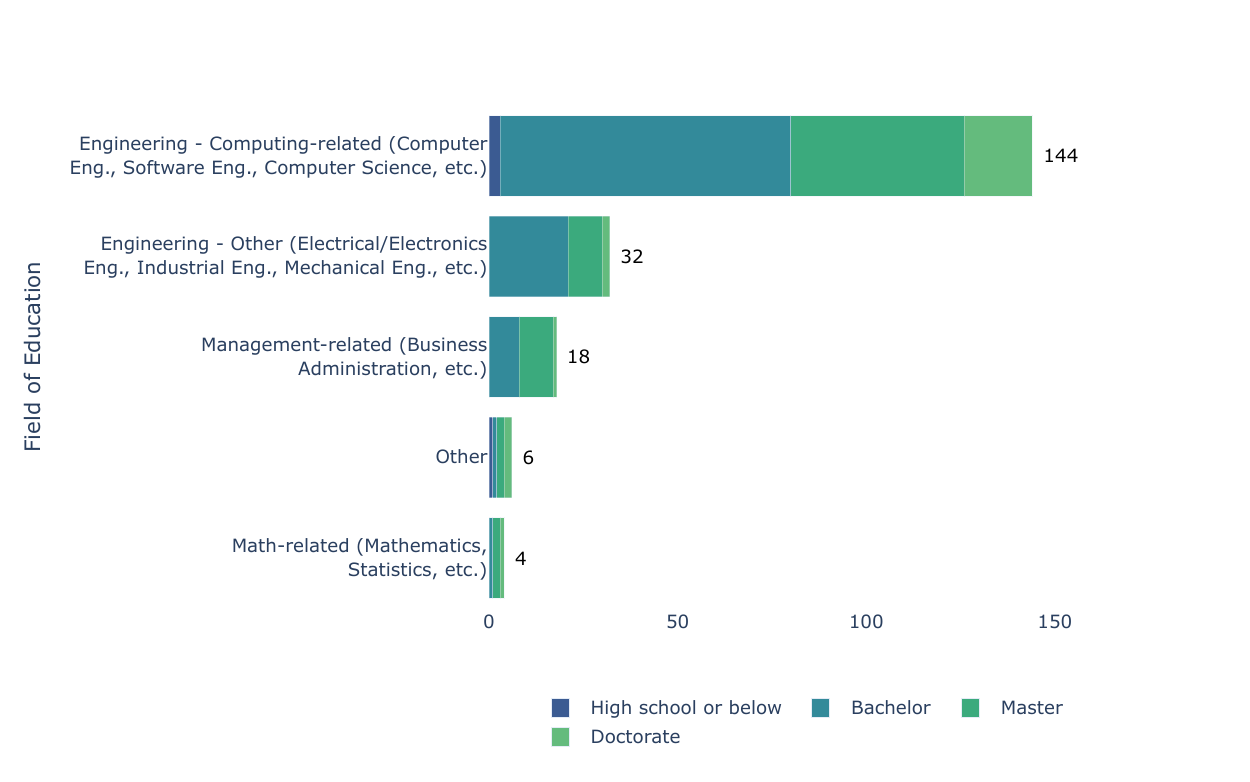}
  \caption{Education field and degree of the respondents (\textit{n} = 204)}
  \label{figure02}
\end{figure}

\textit{Education.} As Figure \ref{figure02} shows, the educational background of most of the respondents is computing-related, e.g., Computer Engineering, Software Engineering, and Computer Science. The next most prevalent fields are other disciplines of engineering, e.g., electrical/electronics, industrial, and mechanical engineering. The rest of the participants have an educational background in management-related, math-related, and other disciplines, such as physics and chemistry. In terms of academic level, bachelor’s degree is the most common qualification at 53\% (\textit{n} = 108) followed by master’s degree at 33\% (\textit{n} = 68). 12\% of the sample held a doctorate degree, which accounted for 24 respondents.

\begin{figure}[h]
  \centering
  \includegraphics[width=\linewidth]{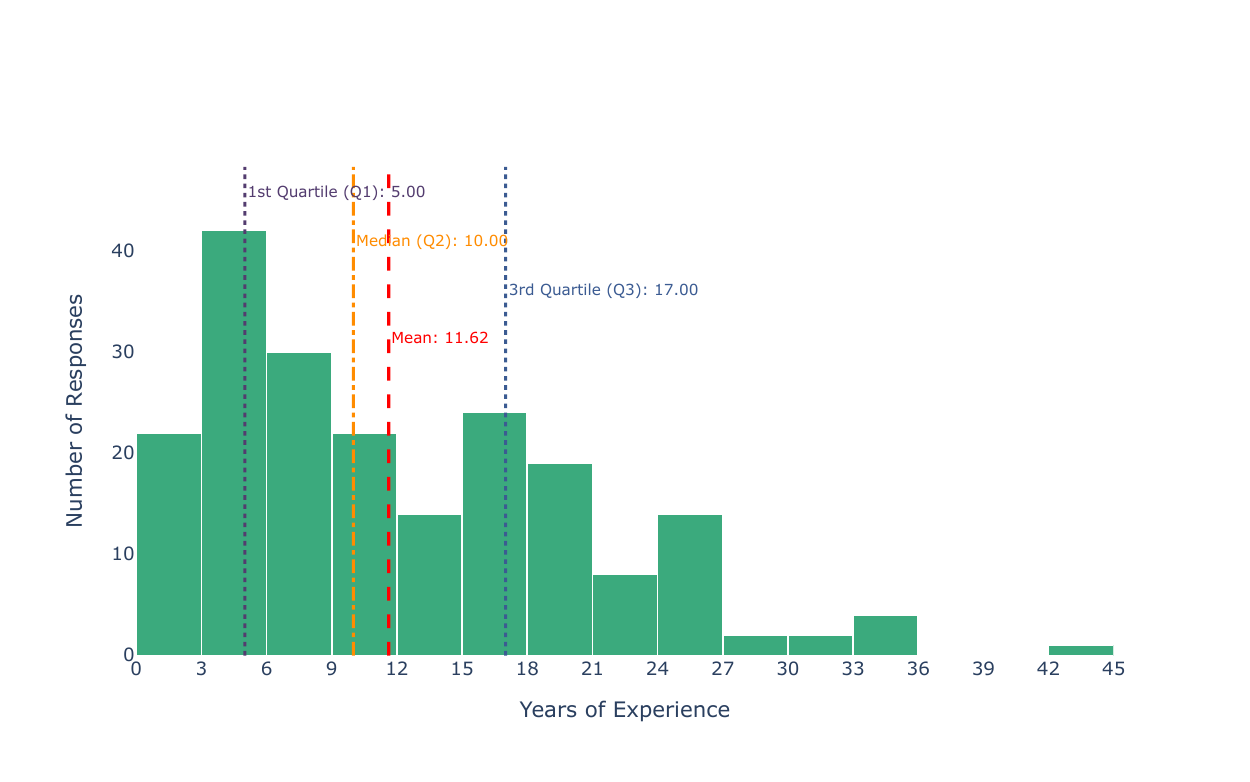}
  \caption{Distribution of the respondents’ experience (\textit{n} = 204)}
  \label{figure03}
\end{figure}

\textit{Experience.} As visualized in the histogram in Figure \ref{figure03}, the distribution of experience is skewed towards shorter careers, indicating a strong presence of early-to-mid career professionals. The average professional experience is 11.6 years, and the median is 10.0 years. The interquartile range reveals that 50\% of the respondents possess experience between 5.0 and 17.0 years.

\begin{figure}[h]
  \centering
  \includegraphics[width=\linewidth]{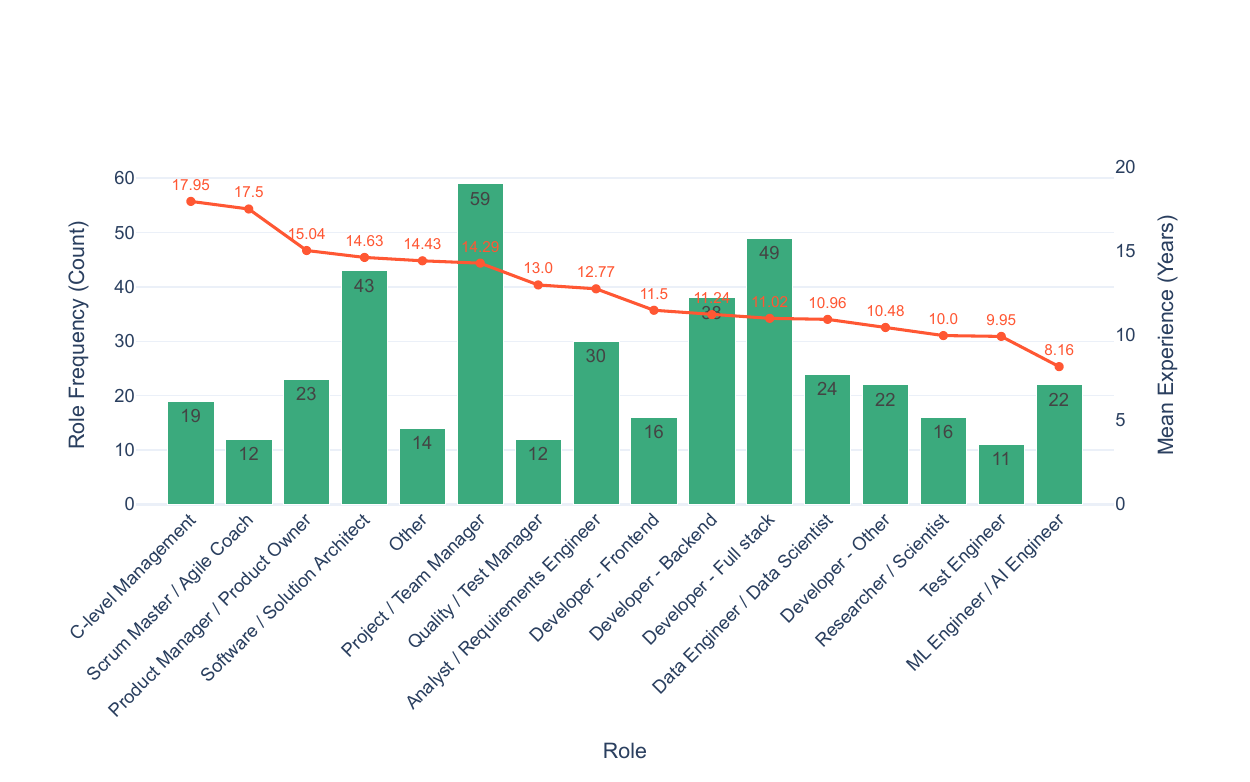}
  \caption{Role frequencies and mean experience per role (\textit{n} = 204)}
  \label{figure04}
\end{figure}

\textit{Role and Experience.} Figure \ref{figure04} shows the role frequencies and the means of experience per role; each respondent may have more than one role. The respondents with strategic and executive roles, e.g., C-level management and project/team manager, are more experienced compared to the respondents with operational roles, such as developer and test engineer. The most frequent role is project/team manager (\textit{n} = 59) with a mean experience of 14.29 years. 19 respondents possess a C-level role and have an average experience of 17.95 years. Core technical roles, such as full stack developer (\textit{n} = 49) and backend developer (\textit{n} = 38) show lower mean experience, i.e., 11.02 and 11.24 years, respectively. Finally, AI/ML engineer (\textit{n} = 22), an emerging role, has the lowest mean experience at 8.16 years.

\begin{figure}[h]
  \centering
  \includegraphics[width=\linewidth]{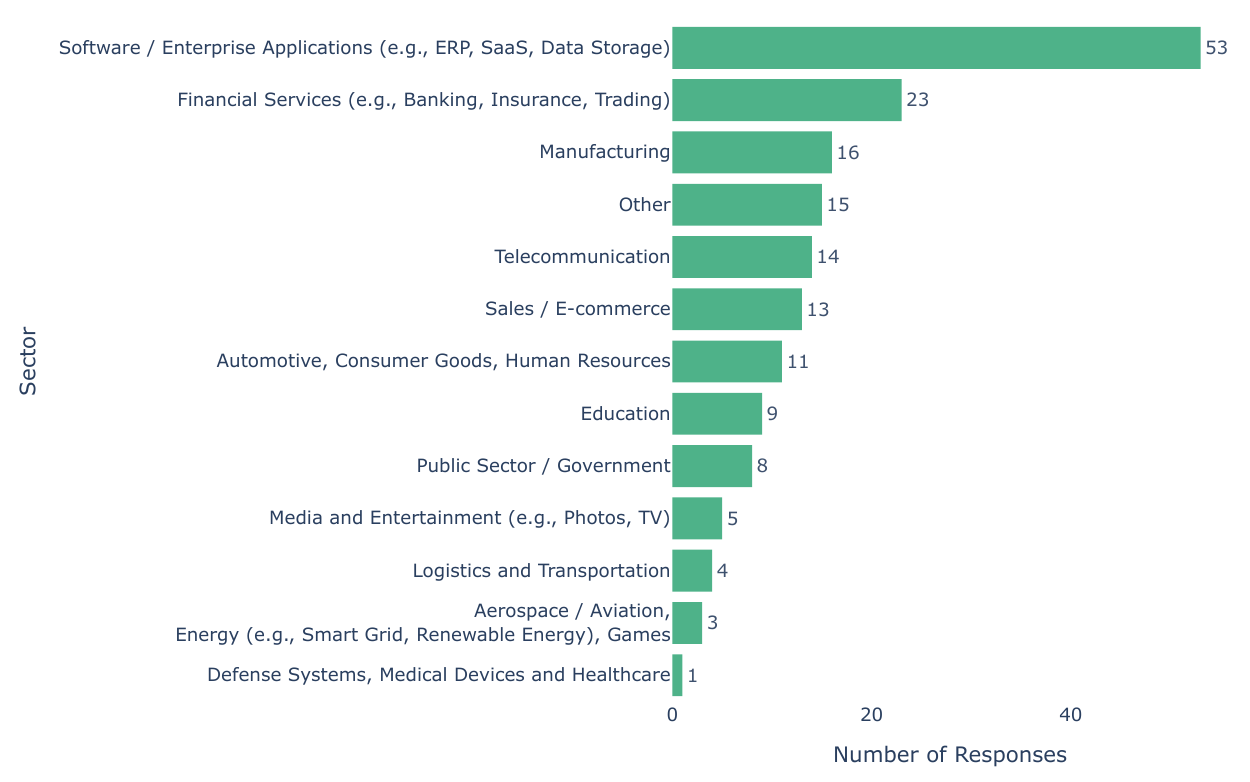}
  \caption{Sector distribution of the respondents (\textit{n} = 204)}
  \label{figure05}
\end{figure}

\textit{Sector.} The distribution of the sectors our respondents are working in is concentrated around software and financial services industries, as shown in Figure \ref{figure05}. 26\% (\textit{n} = 53) and 11\% (\textit{n} = 23) of the respondents work in the software and financial services industries, respectively. The manufacturing, telecommunications and sales/e-commerce sectors follow with 8\% (\textit{n} = 16), 7\% (\textit{n} = 14) and 6\% (\textit{n} = 13). The rest of the respondents are distributed across 15 other sectors, including automotive, consumer goods, and human resources.

\begin{figure}[htbp]
  \centering
  \begin{subfigure}[b]{0.48\textwidth}
    \centering
    \includegraphics[width=\linewidth]{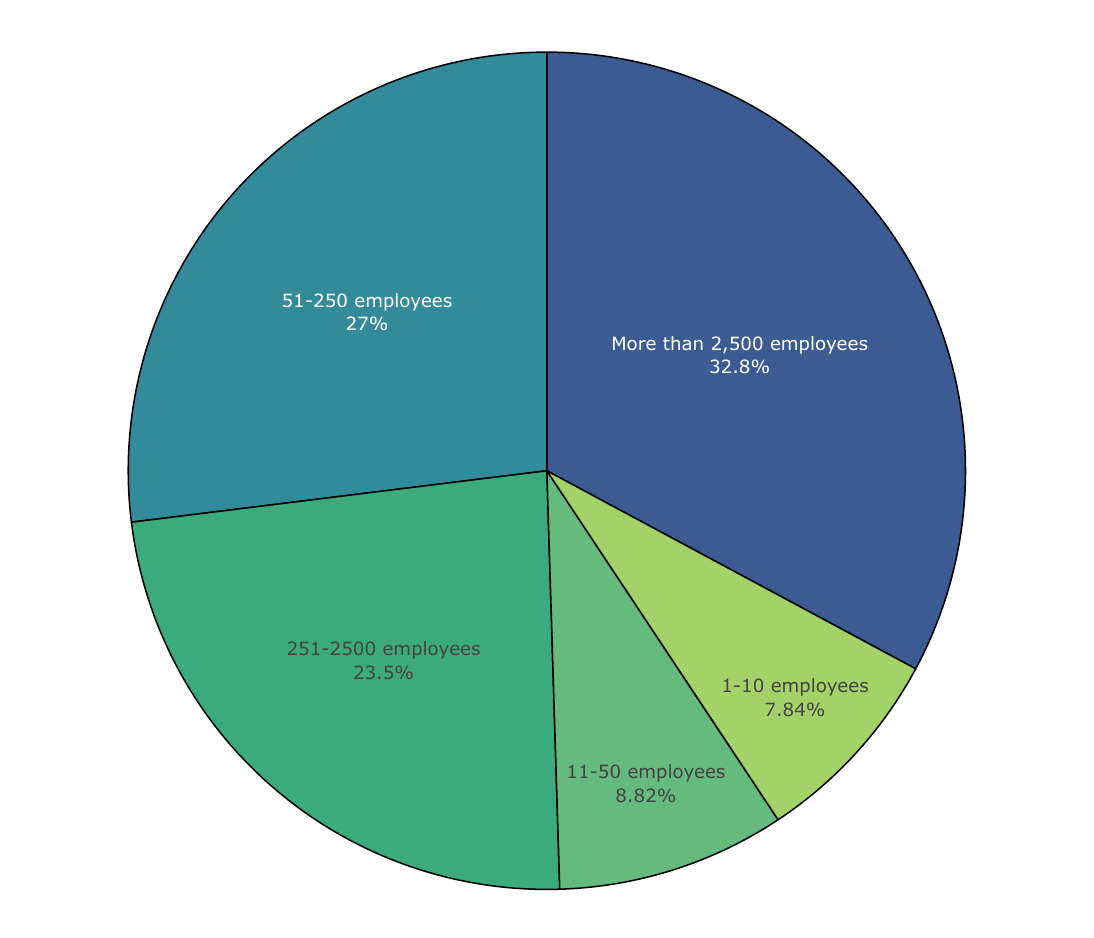}
    \caption{Organization size}
    \label{figure06a}
  \end{subfigure}
  \hfill 
  \begin{subfigure}[b]{0.48\textwidth}
    \centering
    \includegraphics[width=\linewidth]{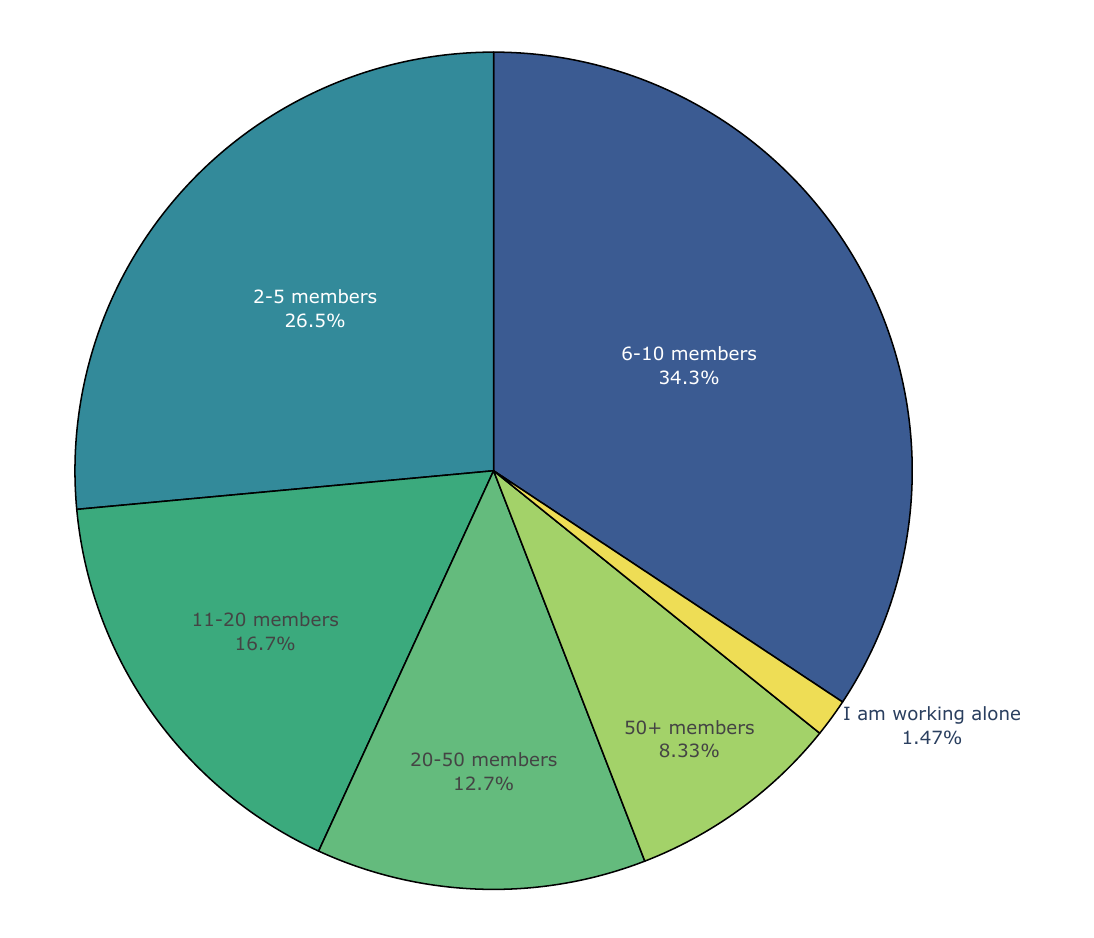}
    \caption{Team size}
    \label{figure06b}
  \end{subfigure}
  
  \caption{Distribution of (a) organization and (b) team sizes (\textit{n} = 204)}
  \label{figure06}
\end{figure}

\textit{Organization and Team Size.} As seen in Figure \ref{figure06}, the most significant segment comprises respondents working at companies with more than 2,500 employees, accounting for one-third of the sample at 32.8\% (\textit{n} = 67). 55 respondents (27\%) are from mid-sized companies (51-250 employees) and 48 (23.5\%) are from larger companies (251-2500 employees). Micro-enterprises with 1-10 and 11-50 employees make up the smallest portions, at 7.8\% (\textit{n} = 16) and 8.8\% (\textit{n} = 18). The composition of team size is highly concentrated in the small to mid-size range. The most prevalent team structure consists of six to 10 members, which accounts for 34.3\% (\textit{n} = 70). 54 respondents (26.5\%) are working in teams made up of two to five members. Larger team structures, i.e., 11-20 members (16.7\%) and 20-50 members (12.7\%) are less frequent. The minority categories consist of very large teams having more than 50 members (8.3\%) and respondents working alone (1.5\%).

\textit{Project Management Approach.} As Figure 8 shows, the most common and self-reported  approach is agile (33.8\%, \textit{n} = 69), followed by somewhat agile (24.5\%, \textit{n} = 50) and hybrid (23.5\%, \textit{n} = 48). Approaches involving agile practices (agile, somewhat agile, hybrid) account for over 82\% of the responses, indicating a strong dominance of iterative and adaptive methods. Plan-driven approaches (plan-driven and somewhat plan-driven) are mostly reported by the respondents in larger scale companies, i.e., with 50+ employees.

\begin{figure}[h]
  \centering
  \includegraphics[width=\linewidth]{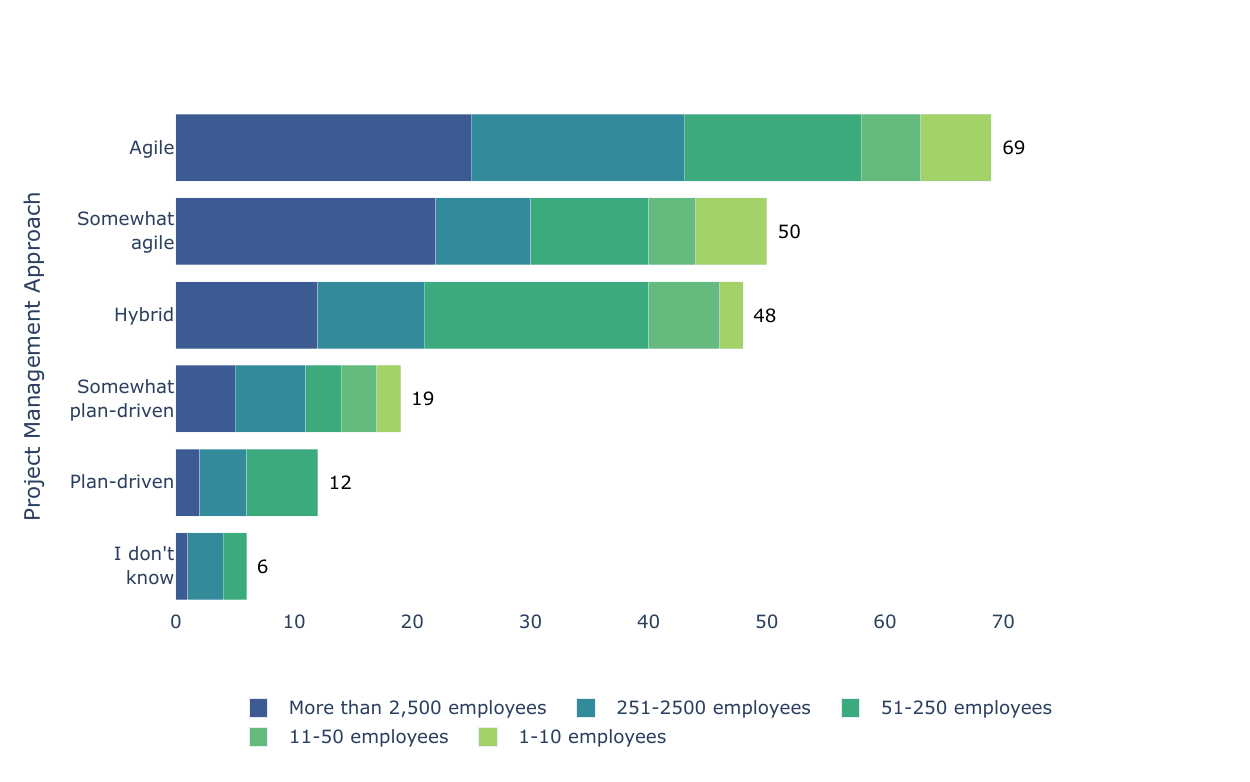}
  \caption{Project Management Approach and split per organization size (\textit{n} = 204)}
  \label{figure07}
\end{figure}

\subsection{RQ1. Status of GenAI tool use in SE}
About 80\% of the respondents (P = 79.44\% [79.27, 79.61]) reported that they utilize GenAI tools in their SE activities. The rest, accounting for one fifth of the respondents (P = 20.56\% [20.39, 20.73]), do not use any GenAI tool for SE-related activities. \added{Across the geographical regions listed in Table \ref{tab:responses_by_regions}, GenAI adoption exhibited notable variation. The highest adoption rate was observed in Latin America and the Caribbean (P = 90.01\% [89.59, 90.43]), followed by the Middle East and North Africa (P = 85.57\% [85.19, 85.95]). Europe and Central Asia (P = 81.58\% [81.26, 81.89]) and North America (P = 81.26\% [80.84, 81.68]) reported comparable adoption levels, both slightly above the overall sample average. In contrast, adoption was lower in East Asia and the Pacific (P = 71.94\% [71.49, 72.39]), South Asia (P = 66.40\% [65.38, 67.42]), and Sub-Saharan Africa (P = 54.87\% [53.77, 55.97]). GenAI adoption also varied across organization sizes. The highest adoption rate was observed among respondents working in organizations with 251–2,500 employees (P = 87.63\% [87.33, 87.93]), followed by organizations with 11–50 employees (P = 83.08\% [82.51, 83.64]). Adoption was somewhat lower among organizations with 51–250 employees (P = 78.43\% [78.08, 78.77]) and those with more than 2,500 employees (P = 77.48\% [77.16, 77.79]). In contrast, the lowest adoption rate was observed among respondents working in organizations with 1–10 employees (P = 62.59\% [61.83, 63.35]).}

\subsubsection{RQ1.1. Reasons behind not using GenAI tools}
\begin{figure}[h]
  \centering
  \includegraphics[width=\linewidth]{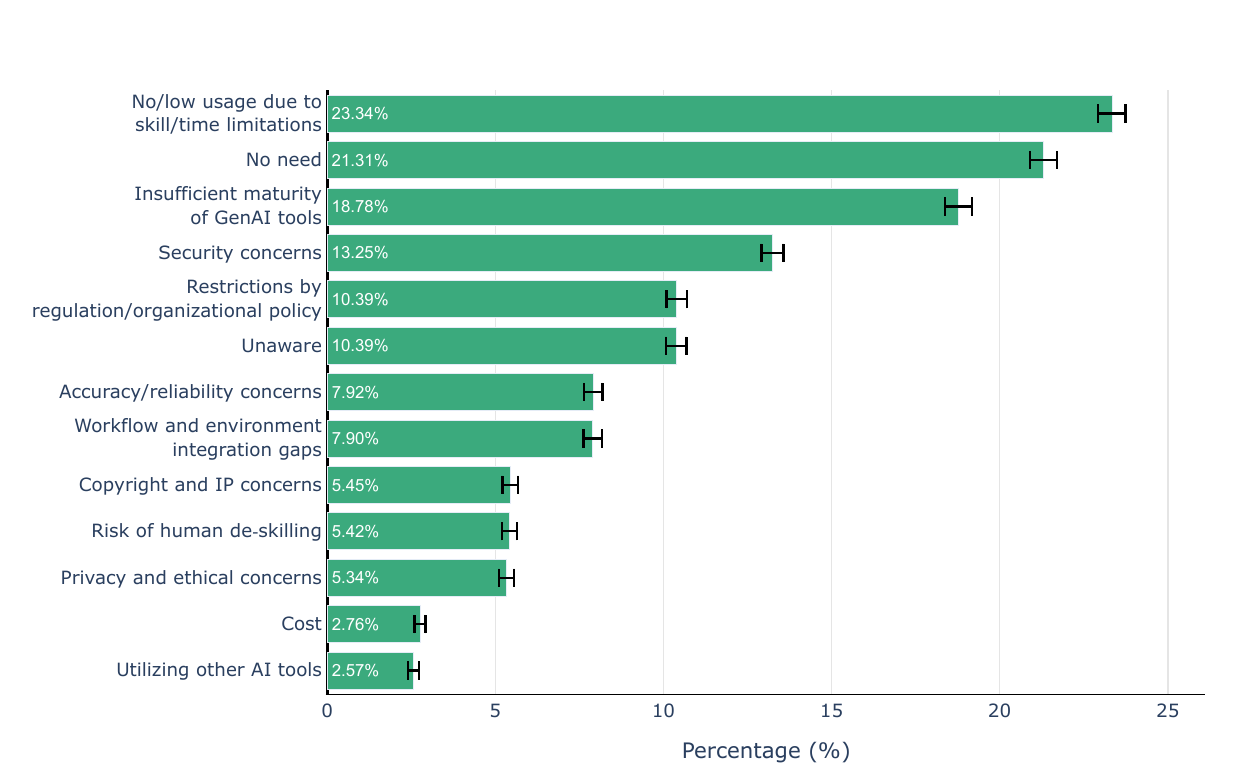}
  \caption{Reasons behind not using GenAI tools (\textit{n} = 38)}
  \label{figure08}
\end{figure}

38 of the respondents, who reported to use no GenAI tool in SE-related tasks, provided one or more reasons. Figure \ref{figure08} shows the reasons behind not using GenAI tools. The most significant barrier is the lack of required skills or time constraints reported by approximately 23\% of the respondents (P = 23.34\% [22.93, 23.74]). 21.31\% [20.91, 21.71] of the respondents indicated that they do not need GenAI tools in their current workflows. 18.78\% do not find the current capabilities of GenAI tools good enough, including a mistrust in the output quality. Security concerns account for 13.25\% [12.92, 13.57]. Restrictions imposed by regulation or organizational policy were a factor for 10.39\% [10.09, 10.69]. A corresponding 10.39\% [10.08, 10.69] of respondents reported being unaware of GenAI tools or their potential applications. Each of the rest of the reasons accounts for less than ten percent and include concerns on accuracy and reliability, workflow and environment integration gaps, concerns on copyright and IP, risk of human de-skilling, concerns on privacy and ethics, high costs, and utilizing other AI tools but not GenAI tools.

\subsubsection{RQ1.2. SE tasks supported by GenAI tools}
\begin{figure}[h]
  \centering
  \includegraphics[width=\linewidth]{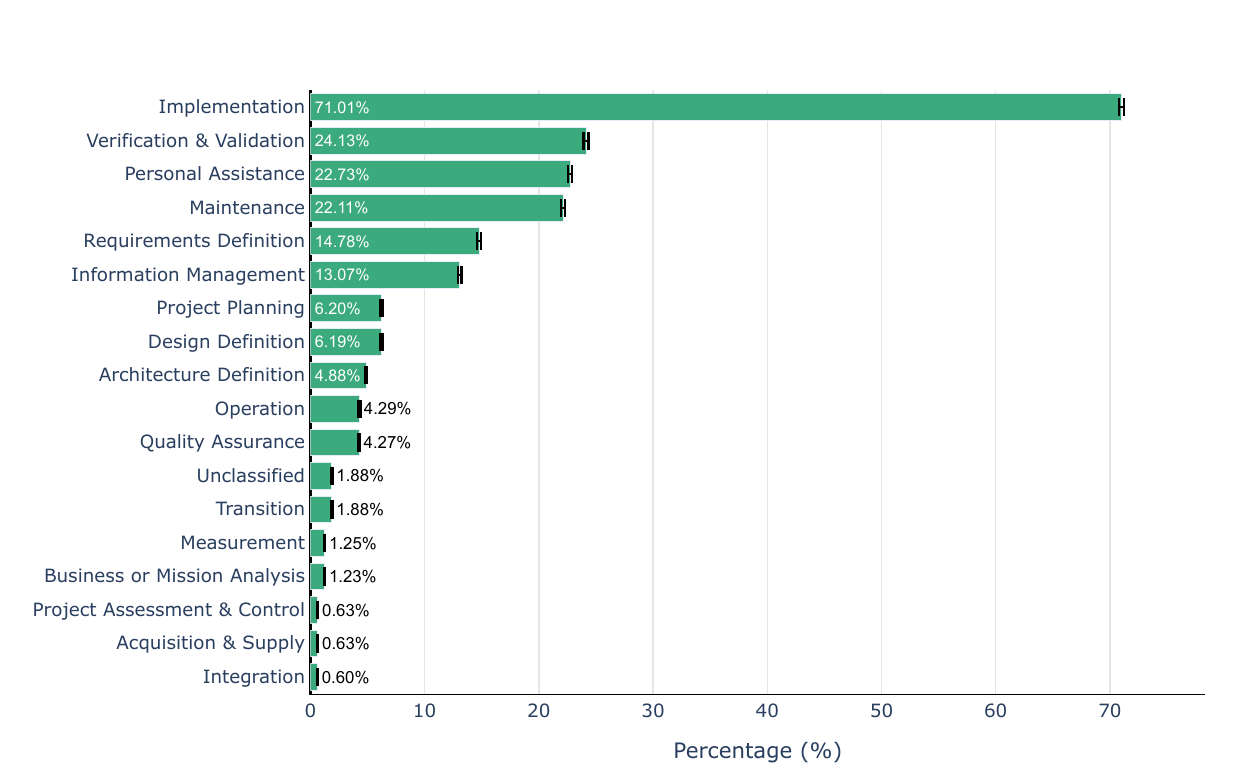}
  \caption{SE tasks supported by GenAI tools (\textit{n} = 162)}
  \label{figure09}
\end{figure}

162 respondents reported one or more SE tasks for which they utilize GenAI tools, averaging approximately two tasks per respondent. We grouped these tasks into higher level groups as shown in Figure \ref{figure09} using the software lifecycle processes in the ISO/IEC/IEEE 12207:2017 standard \cite{ISO12207}. The most frequently reported group of tasks are related to implementation (P = 71.01 [70.79, 71.23]) such as coding, code generation, and code completion. The second most frequent category involves verification and validation activities, including code review and test case and data generation (P = 24.13\% [23.91, 24.34]). More than one fifth of the respondents use GenAI tools for documentation, knowledge search, learning, brainstorming, and problem solving (P = 22.73\% [22.52, 22.93]). Maintenance activities constitute the fourth-largest category (P = 22.11\% [21.91, 22.31]), indicating tasks for improving and fixing existing codebases. Maintenance activities include code understanding, debugging, bug analysis, code optimization, and refactoring.

Following the top four, the respondents reported requirements definition (P = 14.78\% [14.61, 14.95]), information management (P = 13.07\% [12.90, 13.23]), project planning (P = 6.20\% [6.08, 6.32]), and design definition (P = 6.19\% [6.08, 6.31]). Finally, each of the remaining categories accounts for less than five percent, including architecture definition, operation, and quality assurance.

\subsubsection{RQ1.3. GenAI tools used and usage patterns}
\begin{figure}[h]
  \centering
  \includegraphics[width=\linewidth]{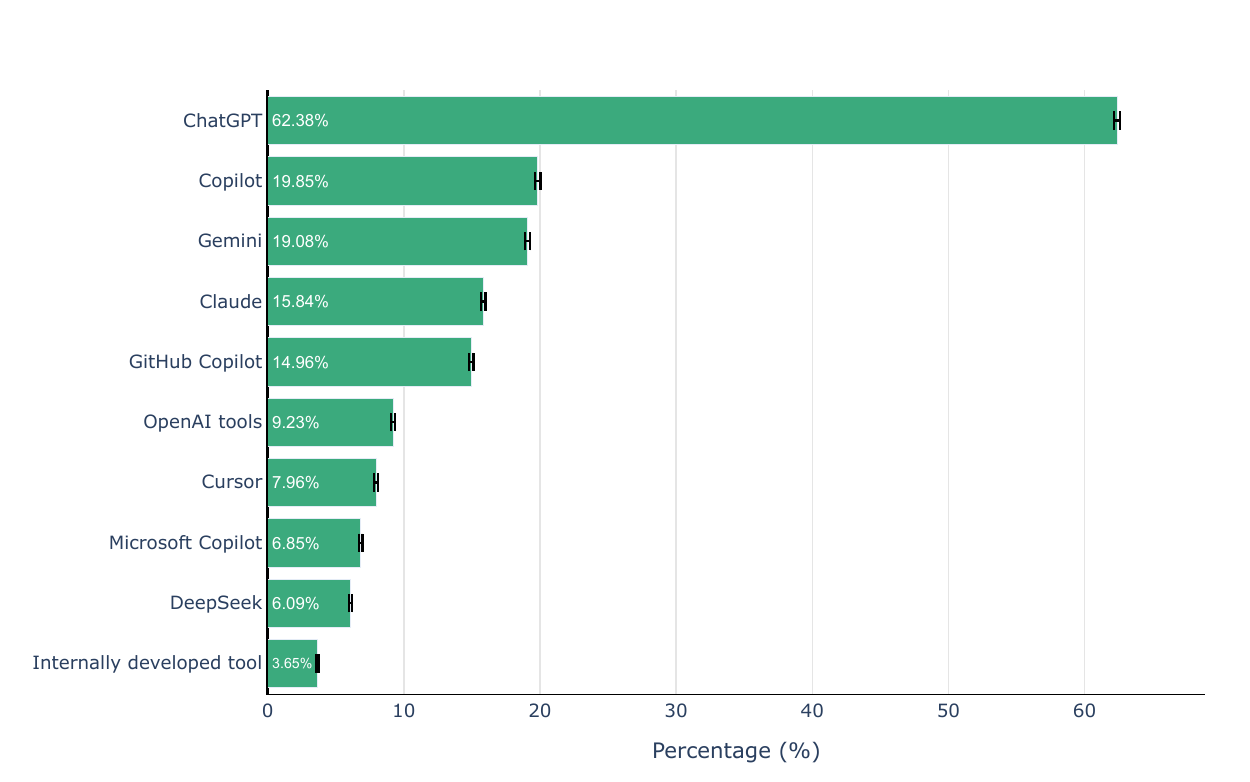}
  \caption{Top 10 GenAI tools used for SE tasks (\textit{n} = 162)}
  \label{figure10}
\end{figure}

162 SE practitioners each reported approximately two GenAI tools on average, totaling up to 48 different mentioned tools. Figure \ref{figure10} shows the top ten most frequently reported GenAI tools. The analysis of the results reveals a strong dominance by general-purpose conversational assistants. ChatGPT (P = 62.38\% [62.14, 62.61]) is by far the most frequently used GenAI tool among SE professionals, followed by Copilot (P = 19.85\% [19.66, 20.03]), Gemini (P = 19.08\% [18.89, 19.28]), and Claude (P = 15.84\% [15.67, 16.01]). While some respondents mentioned only Copilot without providing additional version information, some respondents explicitly mentioned GitHub Copilot (P = 14.96\% [14.79, 15.12]) and Microsoft Copilot (P = 6.85\% [6.72, 6.97]). 9.23\% of the respondents reported use of a tool provided by Open AI. Cursor and DeepSeek have a percentage of 7.96\% and 6.09\%, respectively. 3.65\% reported use of internally developed GenAI tools. Grok (xAI) was reported by 3.60\% of the respondents. 28 of the tools reported have a percentage of less one.

\begin{figure}[h]
  \centering
  \includegraphics[width=0.5\linewidth]{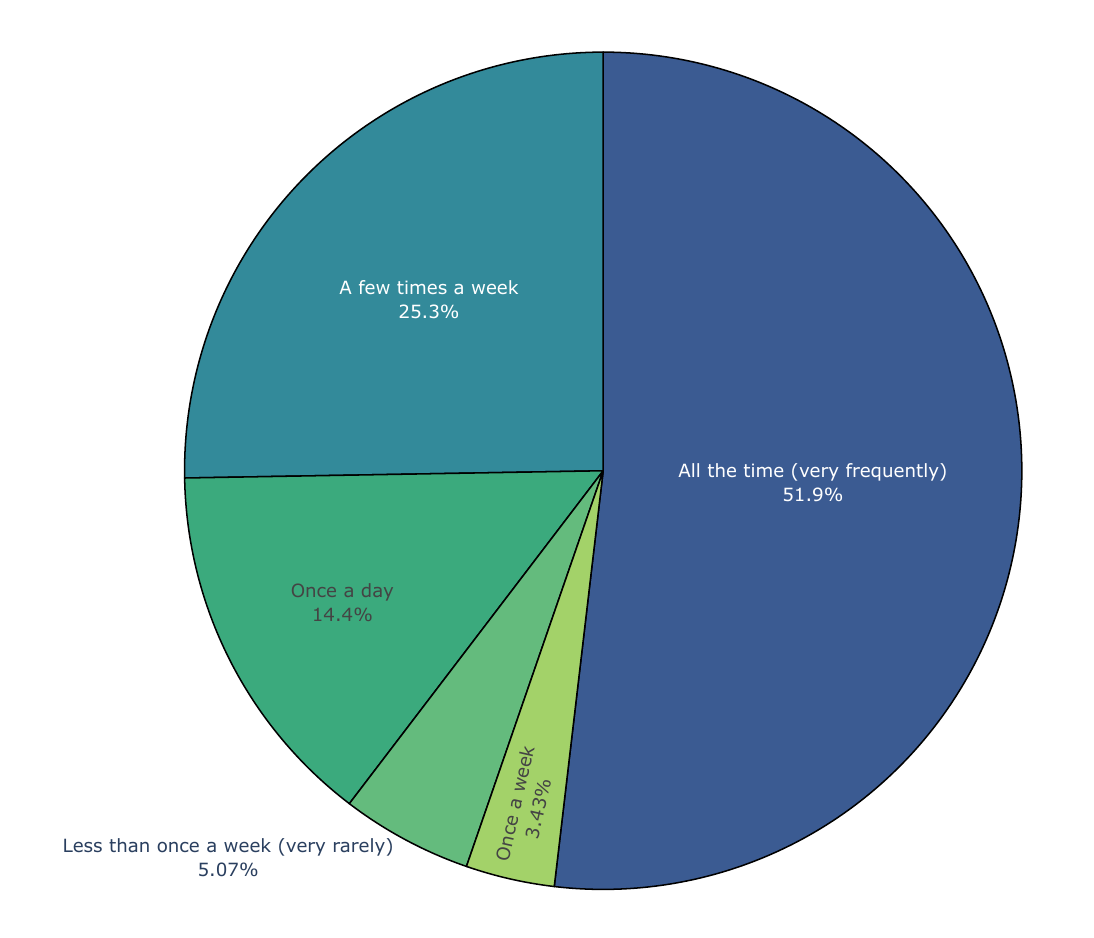}
  \caption{GenAI tool usage patterns (\textit{n} = 162)}
  \label{figure11}
\end{figure}

The self-reported data on GenAI tool usage patterns (Figure \ref{figure11}) indicates a high degree of integration into daily work of SE professionals. 51.9\% [51.63, 52.10] of the respondents reported that they use GenAI tools very frequently. 14.4\% [14.19, 14.52] utilizes GenAI tools once a day. 25.3\% [25.07, 25.48] of the participants reported a few times of usage a week. Only a small fraction (less than 10\%) reported less frequent usage, either once a week or less.

\begin{framed}
\noindent \textit{Summary for RQ1.} \replaced{GenAI tools are widely adopted in SE, with about four out of five respondents reporting active use. Non-use mainly reflects skill and time constraints, limited need, output quality and tool maturity concerns, and security or organizational restrictions. Usage is concentrated in implementation, followed by verification and validation, documentation, learning, and maintenance. General-purpose tools dominate, particularly ChatGPT and Copilot, and most users engage with GenAI daily or near-daily.}{GenAI tools are widely adopted in SE with approximately four out of five respondents reporting active use. Non-use is primarily attributed to skill and time constraints, limited perceived need, concerns about tool maturity and output quality, and security or organizational restrictions. Among users, GenAI tools are predominantly applied to implementation-related activities, followed by verification and validation tasks, documentation and learning-related activities, and software maintenance. Usage is less common in early lifecycle phases such as requirements, design, and project planning. General-purpose conversational GenAI tools dominate the tool landscape, with ChatGPT being the most frequently used, followed by Copilot and other LLM–based assistants. Specialized or internally developed tools are used by a relatively small subset of practitioners. The frequency of use is high, and most users report daily or near-daily interaction with GenAI tools, indicating a strong integration into routine SE workflows.}
\end{framed}

\subsection{RQ2. Benefits and challenges of using GenAI tools for SE tasks}

\subsubsection{RQ2.1. Reported benefits}
\begin{figure}[h]
  \centering
  \includegraphics[width=\linewidth]{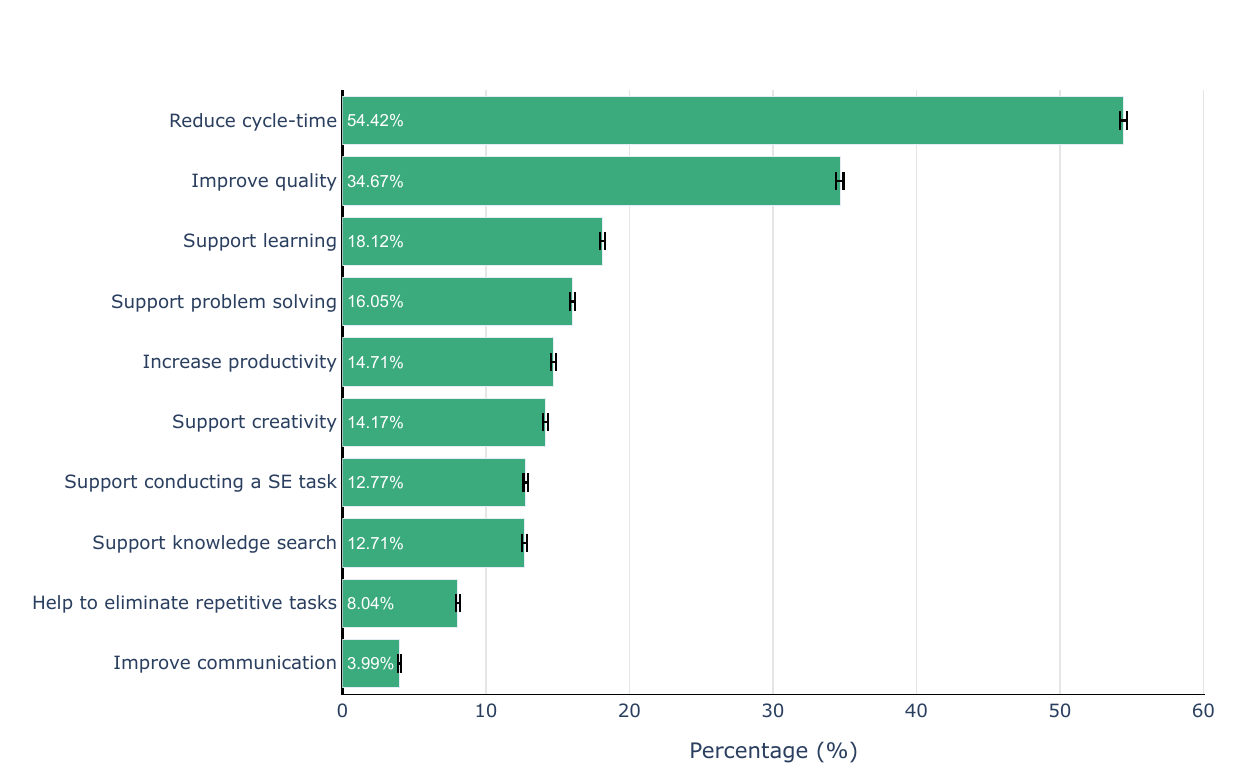}
  \caption{Reported benefits obtained from GenAI tools (\textit{n} = 149)}
  \label{figure12}
\end{figure}

We categorized free text responses into distinct classes of benefits to understand the value propositions of GenAI tools for software engineering. The analysis reveals that the perceived benefits are centered around operational efficiency, quality improvement, and cognitive support to SE practitioners, as seen in Figure \ref{figure12}.

The most prominent perceived benefit, reported by 54.42\% of the participants [54.18, 54.67], is the reduction of cycle time. Participants frequently described GenAI as an enabler for “faster coding, debugging, prototyping” and “time saving”. Quality improvement was the second most cited benefit (P = 34.67\% [34.42, 34.92]). Some respondents noted that GenAI tools help in improving the quality of several SE artifacts, including documentation, code, and query. This indicates that some SE practitioners view GenAI not just as a tool for creating software artifacts faster but also for creating better quality artifacts. Participants reported improvements across multiple dimensions of software artifacts, including code with fewer defects and enhanced efficiency, documentation with clearer content and corrected grammar, and test cases that addressed previously overlooked edge scenarios. 18.12\% [17.93, 18.32] of the respondents find GenAI beneficial for learning. They reported general purpose learning activities, such as learning new concepts, tools, frameworks, as well as project-specific learning, such as codebases and APIs. GenAI tools also support practitioners in solving problems, such as explaining errors, suggesting potential root causes, identifying and resolving bugs (P = 16.05\% [15.87, 16.23]). Closely related to cycle time reduction, productivity increase was also reported as a benefit by 14.71\% [14.54, 14.88] of the respondents. This category is composed of the responses mainly including productivity and efficiency. 14.17\% [13.99, 14.34] of the respondents find GenAI tools very useful in tasks that require creativity, such as brainstorming, generating ideas, and exploring alternatives. Some respondents (P = 12.77\% [12.61, 12.94]) reported getting support for performing a SE task as a benefit. GenAI tools are also seen as alternatives to traditional search engines (P = 12.71\% [12.53, 12.88]). One respondent reported that GenAI tools provide better search results than Google and additionally summarize these results. Less frequent benefits reported are elimination of repetitive tasks and improvement in communication.

\subsubsection{RQ2.2. Perceived productivity change}
\begin{figure}[h]
  \centering
  \includegraphics[width=\linewidth]{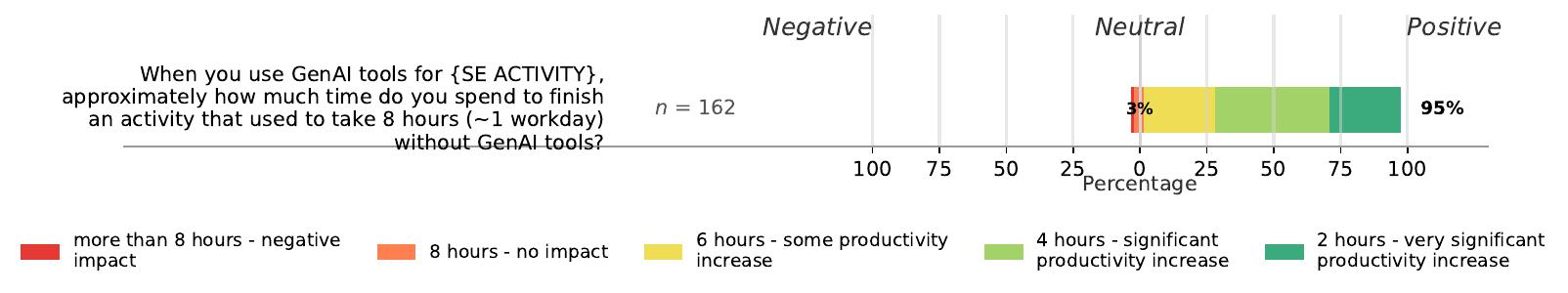}
  \caption{Perceived productivity change (\textit{n} = 162)}
  \label{figure13}
\end{figure}

To capture practitioner perceptions of productivity change, participants were asked to estimate the time they believed would be required to complete an SE task that traditionally requires one workday when assisted by GenAI tools. It should be noted that this question measures subjective time perception rather than objectively recorded task duration, and therefore responses are susceptible to optimism bias and social desirability effects commonly observed in technology adoption surveys \cite{schaupp2010file, ciriello2023yea} and particularly in behavioral studies on AI tool use \cite{yu2026cognitiveoffloadingspeedupillusion}. Figure \ref{figure13} presents the distribution of responses. 43.05\% [42.83, 43.28] of the respondents indicated that they could complete an 8-hour task in a period of four hours (a 50\% reduction), while 26.61\% [26.41, 26.82] reported completing it in two hours (a 75\% reduction). 25.69\% [25.48, 25.89] of the respondents experienced a moderate increase in productivity, reducing the task time to six hours. Only a minority reported no gain (P = 3.49\% [3.40, 3.58]) or a negative impact (P = 1.16\% [1.11, 1.21]). In summary, approximately 95\% of the participants reported a perceived increase in productivity.

\added{When examined by seniority, the perceived productivity impact of GenAI also showed some variation across experience levels. Among junior practitioners (35 respondents with less than or equal to 4 years of experience), 47.03\% [46.54, 47.53] reported reducing an eight-hour task to four hours, while 21.84\% [21.41, 22.28] reported completing it in two hours and 25.31\% [24.87, 25.74] in six hours. Among mid-career practitioners (45 respondents with greater than 4 and less than or equal to 9 years of experience), the four-hour outcome was even more prevalent (P = 52.17\% [51.70, 52.63]), followed by the six-hour (P = 29.35\% [28.96, 29.74]) and two-hour (P = 18.49\% [18.12, 18.85]) outcomes. Notably, no perceived, neutral or negative productivity impact was reported for this group. In contrast, senior practitioners (82 respondents with greater than or equal to 10 years of experience) most frequently reported either completing the task in four hours (P = 36.65\% [36.34, 36.96]) or two hours (P = 30.98\% [30.66, 31.29]), while 25.72\% [25.42, 26.02] reported completing it in six hours. The senior group also exhibited the highest proportion reporting no productivity improvement (P = 5.53\% [5.38, 5.68]), whereas only 1.13\% [1.06, 1.19] reported a negative impact. Overall, perceived productivity gains were prevalent across all three seniority levels; however, mid-career practitioners most frequently reported a 50\% reduction in task completion time, whereas senior practitioners more frequently reported the largest, 75\% reduction compared with the other seniority groups.}

\subsubsection{RQ2.3. Perceived change in SE artifacts’ quality}
\begin{figure}[h]
  \centering
  \includegraphics[width=\linewidth]{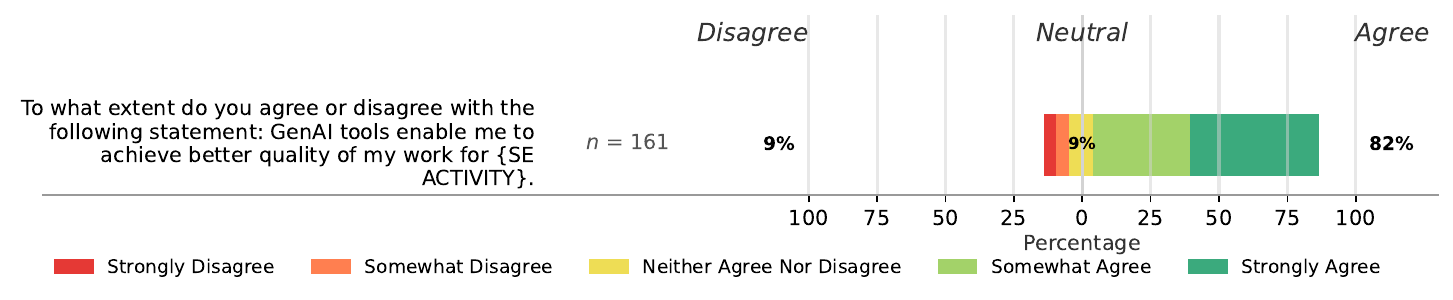}
  \caption{Perceived change in SE artifacts’ quality (\textit{n} = 161)}
  \label{figure14}
\end{figure}

We assessed practitioner perceptions of the impact of GenAI tools on the quality of SE artifacts using a five-point Likert scale. As with the productivity findings above, these responses measure subjective quality perception rather than objectively assessed artifact quality, and should be interpreted accordingly. We assessed the perceived impact of GenAI tools on the quality of SE artifacts using a five-point Likert scale. As Figure \ref{figure14} shows, a combined 82\% of the respondents either “strongly agree” (P = 46.72\% [46.47, 46.97]) or “somewhat agree” (P = 35.21\% [34.99, 35.44]) that GenAI tools improve their quality of work. 8.79\% [8.65, 8.93] of the participants held a neutral position. Only approximately 9\% of the respondents expressed some level of disagreement, i.e., “strongly disagree” and “somewhat disagree” both with 4.64\% [4.54, 4.74].

\added{When examined by seniority, positive perceptions of GenAI's impact on artifact quality were prevalent across all experience groups, although some differences emerged in the distribution of responses. Among junior practitioners (35 respondents with less than or equal to 4 years of experience), 44.27\% [43.73, 44.81] strongly agreed and 36.16\% [35.68, 36.64] somewhat agreed that GenAI tools improve the quality of their work, corresponding to a combined positive assessment of approximately 80.43\%. Among mid-career practitioners (45 respondents with greater than 4 and less than or equal to 9 years of experience), 47.60\% [47.15, 48.05] strongly agreed and 31.17\% [30.78, 31.56] somewhat agreed, resulting in approximately 78.77\% expressing a positive perception. The corresponding proportion was highest among senior practitioners (82 respondents with greater than or equal to 10 years of experience), with 47.22\% [46.86, 47.58] strongly agreeing and 36.99\% [36.66, 37.32] somewhat agreeing, yielding a combined positive assessment of approximately 84.21\%. Negative perceptions were most prevalent among junior practitioners, for whom 11.35\% [11.02, 11.67] strongly disagreed and 8.23\% [7.93, 8.52] somewhat disagreed. In comparison, 8.48\% [8.24, 8.72] of mid-career practitioners somewhat disagreed, with no strongly disagree responses, while only 5.51\% of senior practitioners expressed disagreement, comprising 4.40\% [4.26, 4.54] who strongly disagreed and 1.11\% [1.04, 1.18] who somewhat disagreed. Overall, the results indicate consistently favorable perceptions of GenAI-related quality improvements across seniority levels, with senior practitioners exhibiting the highest combined level of agreement and the lowest combined level of disagreement.}

\subsubsection{RQ2.4. Measurement of size, productivity, and quality} \label{sectionRQ2.4}
\begin{figure}[h]
  \centering
  \includegraphics[width=\linewidth]{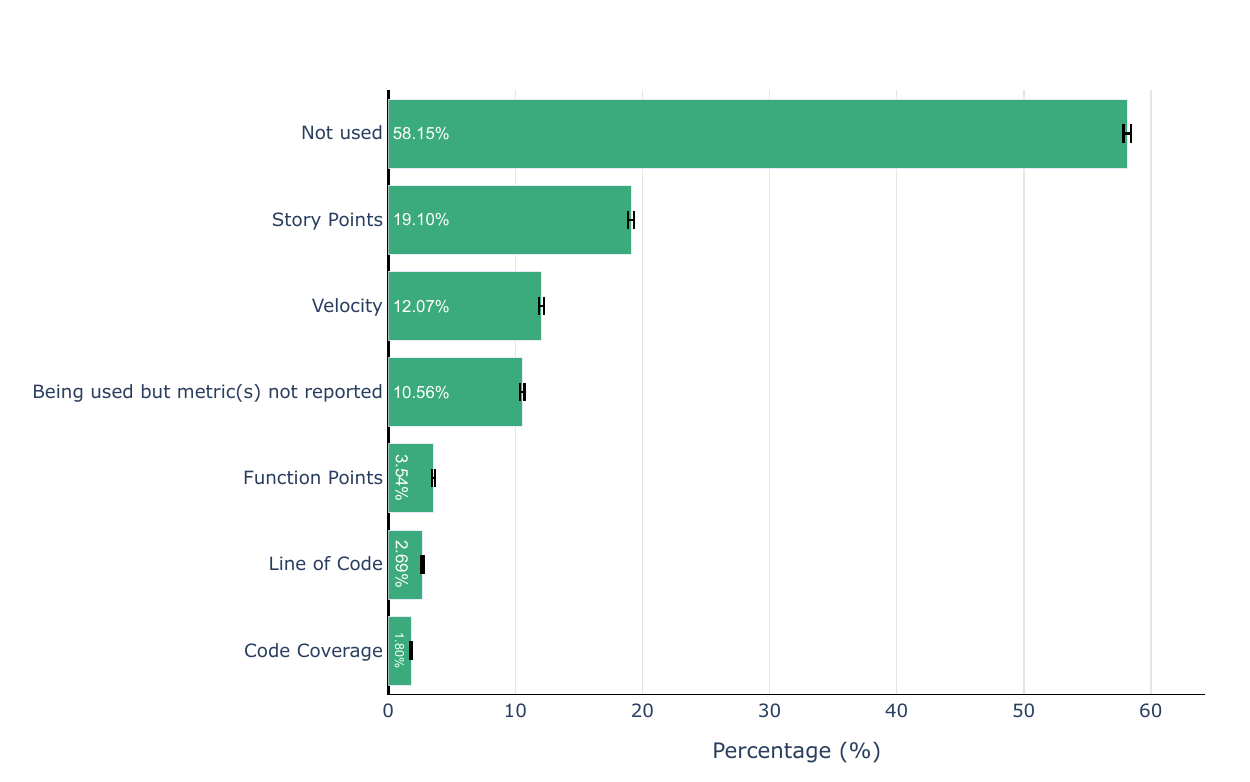}
  \caption{Metrics used for measuring size, productivity, and quality (\textit{n} = 115)}
  \label{figure15}
\end{figure}

We investigated whether SE practitioners utilize objective metrics to quantify size, productivity, and quality in their SE activities. Figure \ref{figure15} shows the responses that have a mean percentage of over one percent. 58.15\% of the respondents [57.85, 58.45] explicitly stated that they do not use any objective metric. Among the respondents who employ objective measures, the most frequently used two metrics are agile development metrics, i.e., story points (P = 19.10\% [18.86, 19.33]) and velocity (P = 12.07\% [11.88, 12.26]). 10.56\% [10.39, 10.73] indicated that they use some metric(s) but did not prefer to disclose which ones. Less common were traditional size metrics, i.e., function points (P = 3.54\% [3.43, 3.65]) and line of code (P = 2.69\% [2.59, 2.78]). Code coverage, a quality metric, was the least cited one having a mean percentage greater than one (P = 1.80\% [1.72, 1.87]). 13 more metrics were reported having less than one mean percentage.

\subsubsection{RQ2.5. Challenges of using GenAI tools for SE tasks}
\begin{figure}[h]
  \centering
  \includegraphics[width=\linewidth]{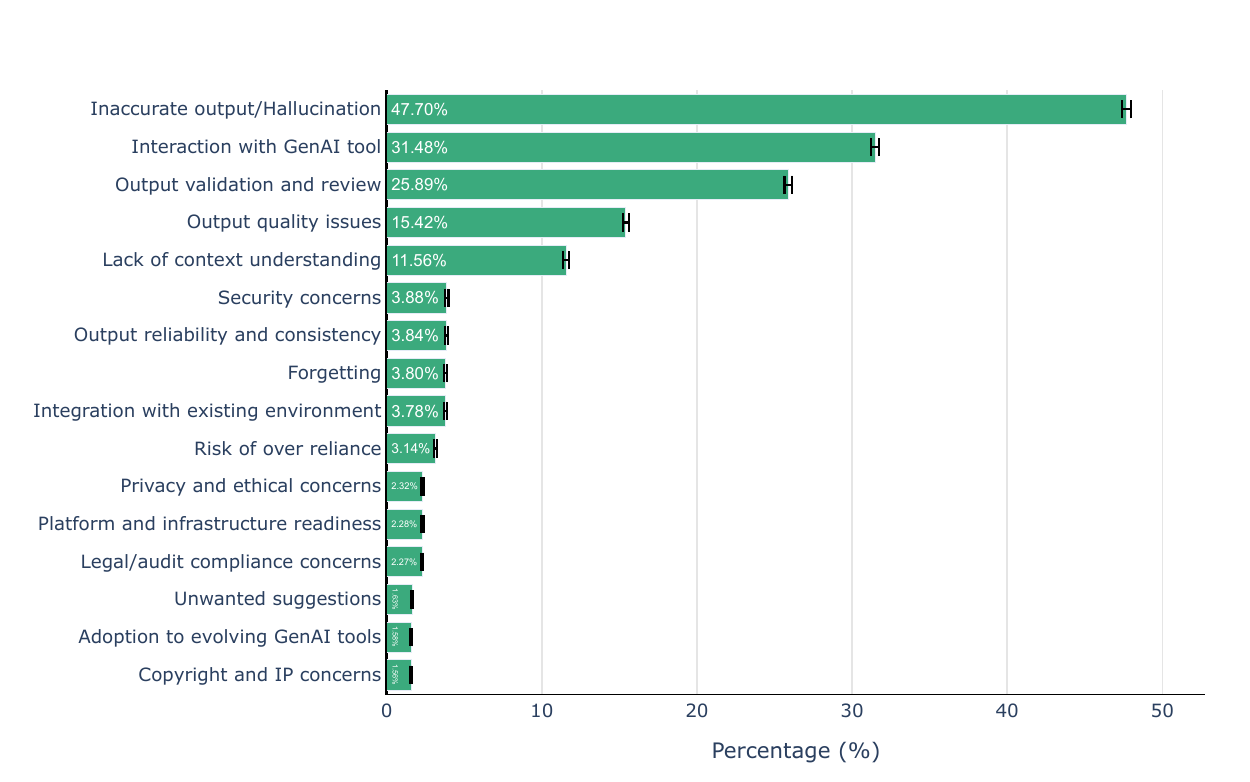}
  \caption{Challenges of using GenAI tools (\textit{n} = 130)}
  \label{figure16}
\end{figure}

To identify the impediments to effective use of GenAI tools, we analyzed participant responses and categorized them into distinct challenge classes. Figure \ref{figure16} presents the frequency analysis for identified challenges having a mean percentage of more than one percent. These challenges were mapped to the characteristics defined in the ISO/IEC 25059:2023 standard (Quality model for AI systems) \cite{ISO25059}. The analysis reveals a variety of challenges that have potential to impact different quality characteristics. Only a small fraction of respondents (P = 4.65\% [4.54, 4.76]) explicitly reported that they do not face any challenge.

\textit{Functional suitability.} The most critical challenges threaten GenAI tools’ capabilities to perform their designated functions accurately. The dominant issue is the generation of incorrect outputs including hallucinations (P = 47.70\% [47.42, 47.98]). Incorrect outputs include defective or insecure code, invalid test cases, and incorrect answers. Hallucinations can be in the form of queries including keywords not present in a database schema, fake references or resources. Apart from inaccurate output, GenAI tools produce suboptimal outputs raising questions on output quality (P = 15.42\% [15.22, 15.61]). Examples of quality issues include low quality code, high-level or abstract responses, and inefficient suggestions. GenAI tools may fail to understand the context (P = 11.56\% [11.38, 11.73]). Context understanding in the SE domain can be making sense of polyrepo codebases, project specific requirements, organizational or regulatory standards. Forgetting, which determines how much of a conversation a GenAI tool can carry out without overlooking details, can lead to the reintroduction of previous defects when fixing another issue or losing track of the constraints provided earlier (P = 3.80\% [3.70, 3.91]). Sometimes code assistants provide suggestions when users do not need (P = 1.63\% [1.57, 1.70]).

\textit{Usability.} Users, especially novices, struggle to prompt GenAI tools properly (P = 31.48\% [31.23, 31.74]). Practitioners frequently reported a “trial and error” dynamic, noting that output quality is heavily dependent on their ability to craft precise, highly detailed prompts. Without providing exhaustive context or splitting prompts into granular instructions, tools often misunderstand the request or hallucinate solutions that ignore project-specific constraints. Furthermore, practitioners find it challenging to adopt rapidly evolving GenAI tools (P = 1.58\% [1.51, 1.65]) and select proper tools for specific SE tasks (P = 0.79\% [0.74, 0.84]). Three participants mentioned that a critical challenge yet to be fully addressed is the lack of transparency and explainability in how GenAI models produce code and suggestions, which consequently erodes developer trust. One participant stated that “it is critical to establish clear mechanisms for auditing, validating, and explaining AI-generated code or decisions – especially in safety-critical or highly regulated domains.”

\textit{Reliability.} The unpredictability of GenAI tools lead to reliability issues. The difficulty of reviewing and validating GenAI output adds a significant burden to practitioners (P = 25.89\% [25.65, 26.13]). Participants emphasized that “code provided tends to be very generic, needs a lot of oversight,” shifting effort from implementation to validation. This validation overhead requires not only time allocation but also being experienced to prevent introducing critical defects in production code. Users reported that the same prompt often yields different results or that performance degrades as “the project gets bigger,” making the tools’ output unreliable and inconsistent (P = 3.84\% [3.74, 3.94]).

\textit{Security.} Participants cited their concerns on security (P = 3.88\% [3.77, 3.98]) and data privacy (P = 2.32\% [2.24, 2.40]). Eleven more participants, who did not report any security concerns in response to the question on challenges, mentioned their concerns under additional important issues, i.e., 30th question. Practitioners raised concerns on the risk of inadvertently sharing proprietary code, business logic, or sensitive data when using cloud-based GenAI tools. This data leakage may lead to intellectual property infringement and legal incompliance.

\textit{Compatibility and Portability.} The integration of GenAI tools into established IDEs, pipelines, or legacy codebases is not straightforward (P = 3.78\% [3.67, 3.88]). A small fraction of respondents reported platform and infrastructure issues related to interoperability and adaptability (P = 2.28\% [2.20, 2.36]).

\textit{Freedom from risk.} One of the social risks of GenAI tools is the risk of overreliance (P = 3.14\% [3.04, 3.23]). Seven participants mentioned the potential negative impact of GenAI on human skills. A critical concern is the potential for overreliance on GenAI tools to significantly weaken fundamental coding, debugging, and critical thinking skills, particularly among novice practitioners. A minority of participants raised concerns on legal and audit compliance (P = 2.27\% [2.19, 2.35]) as well as copyright and IP (P = 1.56\% [1.49, 1.63]). Another seven participants reported their concerns on ethical topics, IP, and legal compliance. Two participants proposed the establishment of guidelines by governments and organizations, which may help to eliminate these concerns. Only one participant raised an environmental risk under additional important issues: “Training and running large foundation models is extremely computationally intensive and has a significant carbon footprint.”

\textit{Other.} Other concerns are prices of GenAI tools, resistance of team members in tool adoption, and access limitations imposed by companies to GenAI tools.

\begin{framed}
\noindent \textit{Summary for RQ2.} \replaced{GenAI tools are widely perceived to improve productivity, quality, and knowledge work in SE. However, these benefits are largely self-reported, as most practitioners use no objective productivity or quality metrics. Key challenges include unreliable outputs, effective prompting, and validation effort, alongside concerns about security, tool integration, and overreliance.}{GenAI tools are widely perceived as beneficial in SE, primarily by increasing productivity, improving quality, and supporting knowledge work. However, these findings are grounded in self-reported perceptions rather than objective measurement: the majority of practitioners report using no objective productivity or quality metrics, which means the strong positive perceptions documented here reflect practitioner experience and sentiment rather than independently validated performance gains. Practitioners also report notable challenges, particularly related to incorrect or unreliable outputs, difficulties in effective prompting, and the effort required to validate GenAI-generated results. Additional concerns include security, integration with existing tools, and the risk of overreliance.}
\end{framed}

\subsection{RQ3. Status of Institutionalization of GenAI tools and techniques in SE}
To understand the integration of Generative AI (GenAI) into organizational practice, we asked whether organizations provide any support for the use of GenAI tools in SE activities. Nearly two-thirds of the respondents (P = 64.57\% [64.37, 64.78]) provide support for GenAI tool use. Conversely, 35.43\% [35.22, 35.63] of respondents reported that their organization does not provide any support. \added{A deeper cross-analysis reveals a strong alignment between tool adoption and the presence of organizational support: Among the 162 respondents who actively use GenAI tools in their workflows, 73.82\% [73.60, 74.04] state that their organizations provide targeted support. In contrast, among the 42 respondents who do not utilize GenAI tools, 71.23\% [70.78, 71.68] report that their organizations do not provide any kind of support.}

\added{Organizational support also varies substantially with organization size. Respondents working in organizations with more than 2,500 employees reported the highest level of support for GenAI tool use (P = 83.50\% [83.22, 83.78]), followed by those in organizations with 251-2,500 employees (P = 66.66\% [66.24, 67.08]). Support was less prevalent in organizations with 51-250 employees (P = 56.20\% [55.78, 56.61]) and was nearly evenly divided among organizations with 11-50 employees, where 49.67\% [48.92, 50.42] reported receiving organizational support. The lowest level of support was observed in organizations with 1-10 employees (P = 24.70\% [24.01, 25.39]), where approximately three-quarters of respondents reported receiving no organizational support (P = 75.30\% [74.61, 75.99]). Overall, these results reveal a clear pattern in which reported organizational support for GenAI tends to be greater in larger organizations, with particularly pronounced differences between the smallest and largest organizations.}

\begin{figure}[h]
  \centering
  \includegraphics[width=\linewidth]{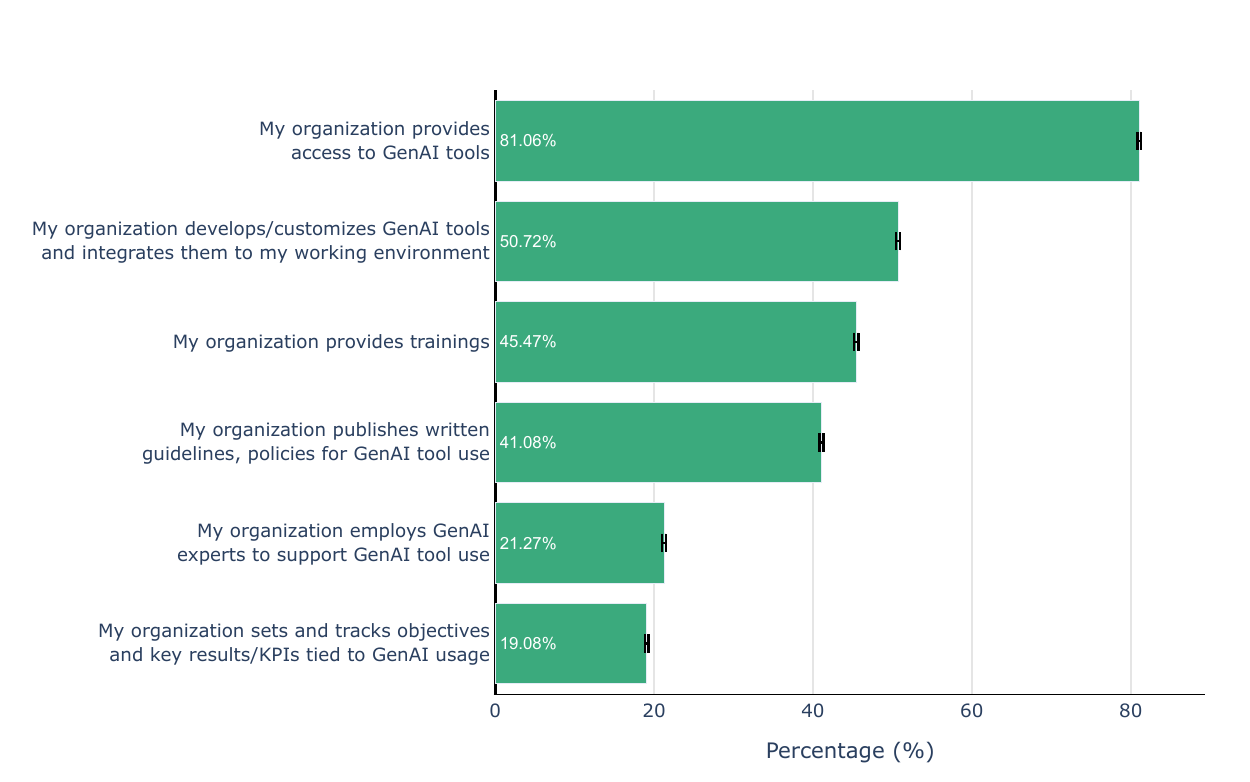}
  \caption{How organizations support GenAI use (\textit{n} = 132)}
  \label{figure17}
\end{figure}

We also investigated the specific approaches organizations are employing to facilitate the use of GenAI tools for SE activities. As Figure \ref{figure17} shows, the most prevalent form of support is the provision of direct access to tools (P = 81.06\% [80.85, 81.27]). Over half of the organizations are moving beyond providing access by actively developing and customizing GenAI tools and integrating them into practitioners’ working environment (P = 50.72\% [50.44, 50.99]). While providing the tools is the top priority, upskilling and governance are also significant organizational focuses. 45.47\% [45.20, 45.74] of organizations provide training to their employees. 41.08\% [40.82, 41.33] have published policies and guidelines, reflecting an active effort to manage the risks associated with GenAI. Only 21.27\% [21.04, 21.49] of organizations employ dedicated GenAI experts to support tool use. The least reported form of organizational support is setting and tracking objectives and KPIs tied to GenAI usage (P = 19.08\% [18.87, 19.29]).

\added{When examined by organization size, several notable differences in organizational support emerged. Direct access to GenAI tools remained highly prevalent across organization sizes and reached 100\% among organizations with 11–50 employees. However, larger organizations demonstrated a markedly broader range of support mechanisms beyond merely providing access. In particular, the development and customization of GenAI tools increased consistently with organization size, rising from 0\% among organizations with 1–10 employees to 33.04\%, 35.64\%, 47.22\%, and finally 67.49\% among organizations with more than 2,500 employees. Training also displayed a strong overall upward trend with organization size, increasing from 24.14\% in the smallest organizations to 58.93\% in the largest, despite a minor decline between the 11–50 and 51–250 employee categories. Moreover, organizations with more than 2,500 employees exhibited the highest levels of support through dedicated GenAI experts (P = 33.71\% [33.31, 34.12]) and through setting and tracking GenAI-related objectives and KPIs (P = 30.39\% [30.02, 30.77]). In contrast, the smallest organizations reported no development or customization of GenAI tools, employment of dedicated GenAI experts, or tracking of GenAI-related objectives and KPIs. Overall, these findings suggest that although providing access to GenAI tools is common across organizations of different sizes, increasing organization size is associated with a shift toward a more comprehensive support infrastructure, particularly through greater investment in developing and customizing GenAI tools and employee training.}

\begin{framed}
\noindent \textit{Summary for RQ3.} \replaced{Organizational support for GenAI use in SE is becoming common but remains inconsistent. About two-thirds of respondents receive some support, mainly through tool access, customization and integration of GenAI tools into existing environments, training, and usage policies. However, dedicated GenAI experts and formal objectives or performance indicators remain less common.}{Institutionalization of the use of GenAI tools in SE is becoming common but remains inconsistent across organizations. Approximately two-thirds of respondents report receiving some form of organizational support, while over one-third operate without any formal support mechanisms. Among supporting organizations, the dominant strategy is providing direct access to GenAI tools. A substantial proportion of organizations go further by developing, customizing, and integrating GenAI tools into existing environments. In addition to tool access, many organizations invest in employee training and establish policies or guidelines to govern GenAI use, reflecting growing awareness of skill and risk management needs. However, fewer organizations allocate dedicated GenAI experts or define formal objectives and performance indicators related to GenAI adoption.}
\end{framed}

\subsection{RQ4. Expected impacts of GenAI tools on SE Community}
The findings reported in this section capture subjective practitioner expectations and perceptions about possible futures rather than observed or measurable outcomes. Given the rapid and unpredictable evolution of GenAI technologies and labor markets, these responses should be interpreted as attitudinal data reflecting how SE professionals currently anticipate change, not as evidence-based forecasts of what will occur.

\begin{figure}[h]
  \centering
  \includegraphics[width=\linewidth]{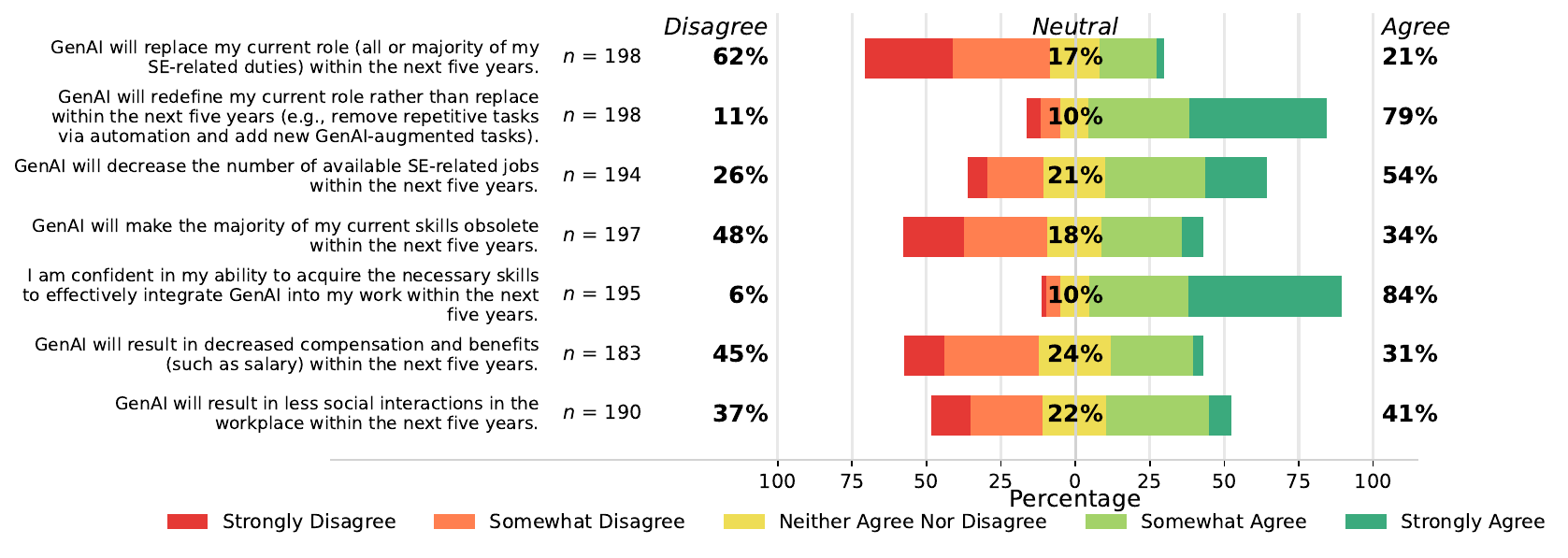}
  \caption{Statements about the potential social impact of GenAI tools (\textit{n} = 183 – 198)}
  \label{figure18}
\end{figure}

Practitioners in our sample exhibit a high degree of confidence and a strong expectation that GenAI will redefine, rather than eliminate, their current roles, as seen in Figure \ref{figure18}. While 62\% of the respondents disagree that GenAI tools will replace their role within five years, 21\% perceive a threat of complete role automation. The majority, accounting for 79\%, expect that GenAI will redefine their role rather than replace. 11\% disagree with this expectation. Despite the confidence in role preservation, there is a moderate concern about the overall job market. 54\% of the participants anticipate that the job market may shrink within the next five years even if their individual roles remain secure, likely due to increased efficiency.

Regarding skill transformation, respondents express considerable confidence in their ability to adapt: 84\% believe that they can acquire the skills required to effectively integrate GenAI into their work, and 48\% disagree that their current skills will become obsolete within the next five years. On economic impacts, 45\% do not expect negative effects on compensation, while 31\% express that efficiency gains provided by GenAI may lead to downward pressure on salaries.

The anticipated impact on workplace social dynamics is viewed as largely neutral, with a slight tendency towards disagreement. The responses were closely split, with 41\% agreeing and 37\% disagreeing, indicating no strong consensus on whether increased GenAI usage will isolate employees.

\begin{framed}
\noindent \textit{Summary for RQ4.} \replaced{Practitioners largely expect GenAI to transform rather than replace SE roles, while remaining confident in their ability to adapt and acquire new skills. However, many anticipate a smaller job market and some expect downward pressure on salaries. These findings reflect practitioner expectations rather than evidence-based forecasts and should be interpreted with appropriate uncertainty.}{The findings summarized below reflect practitioner perceptions and expectations about possible futures, not observable outcomes or evidence-based forecasts. Given the pace of change in GenAI capabilities and labor markets, these attitudinal signals should be interpreted with appropriate uncertainty. SE practitioners in our sample largely expect GenAI tools to transform, rather than replace, their roles in the near future. Most respondents express confidence that their roles will be redefined through collaboration with GenAI, although a minority anticipate full role automation within five years. Despite this confidence at the individual level, practitioners express moderate concern about broader labor market effects, with over half expecting a contraction in the job market driven by efficiency gains. Respondents also report high confidence in their ability to adapt to GenAI-driven changes, with a large majority believing they can acquire the necessary skills and rejecting the notion that their current skill sets will become obsolete. Perceptions of economic impact are mixed: while many do not foresee negative effects on compensation, a substantial proportion anticipate potential downward pressure on salaries. Finally, the anticipated impact of GenAI on workplace social dynamics is largely neutral, with no clear consensus on whether increased GenAI use will lead to employee isolation.}
\end{framed}

\section{Discussion} \label{discussion}
Based on our research questions, we group the discussion into four subsections, i.e., the status of GenAI tool use, benefits and challenges, institutionalization of tools and techniques, and potential impact on the SE community and discuss therein our results also in relation to existing evidence.

\textit{Status of GenAI use.} Our findings indicate a remarkably high penetration of GenAI tools, with approximately 80\% of respondents reporting their use in SE activities. This figure aligns closely with the StackOverflow Developer Survey survey \cite{StackOverflow2025}, which reports that 84\% of respondents are currently using or planning to use AI tools. However, our data suggests a more accelerated adoption than estimated by Capgemini Research Institute \cite{Capgemini2024}, which places current adoption at 46\%, projecting it to reach 85\% only by 2026.

Regarding the usage patterns, our results show that 52\% of respondents use GenAI tools “very frequently”. This is very close to the StackOverflow Developer Survey’s \cite{StackOverflow2025} finding, where 51\% of professional developers reported daily use. This suggests that for adopters, GenAI has transitioned from an experimental novelty to a daily operational necessity.

In terms of tooling, a strong dominance of general-purpose models is evident. We found ChatGPT to be the most frequently used tool (62\%), followed by Copilot (20\%). This corroborates MIT’s “State of GenAI in Business” report’s observation that generic tools like ChatGPT are widely used due to their flexibility and immediate utility, whereas custom solutions often stall due to integration complexity \cite{Challapally2025}.

Our survey identified implementation as the primary use case (71\%), followed by “verification \& validation” (24\%). This supports Capgemini’s finding that “coding assistance” is the leading use case \cite{Capgemini2024} and McKinsey’s assessment that value is derived from generating code drafts and refactoring \cite{McKinsey2023}. Similarly, OpenAI reports that users with engineering roles utilize ChatGPT mostly for programming tasks \cite{OpenAI2025}. On the other hand, the StackOverflow Developer Survey \cite{StackOverflow2025} ranks SE activities slightly differently, listing search for answers (54\%) as the top activity, significantly higher than writing code (17\%). This nuance indicates that while developers label their activity as implementation, the actual cognitive mechanism often involves using GenAI as an advanced search engine to solve problems before code is committed – a behavior also supported by our finding that 23\% use it for knowledge search and problem solving.

When analyzing the reasons for non-adoption, our respondents cited the “lack of required skills or time constraints” (23\%) as the primary barrier, followed by a perceived “no need” (21\%). While mistrust in output quality was present (19\%), it was not the leading factor. Conversely, other studies portray trust and accuracy as the dominant inhibitors. The StackOverflow Developer Survey \cite{StackOverflow2025} reports that more developers actively distrust (46\%) AI accuracy than trust it, with frustration stemming from solutions that are “nearly accurate but requiring refinement”. Similarly, Liang et al. \cite{Liang2024} found that the primary motivations for non-use were that models failed to meet functional requirements (54\%) or were difficult to control (48\%). This suggests a dichotomy: non-users in our sample feel unprepared to use the tools (lack of required skills or time constraints), whereas the broader market feels the tools are unprepared for them (mistrust in AI tools and their output).

\textit{Benefits and challenges.} Our findings indicate that the primary perceived value of GenAI tools in SE lies in their ability to reduce cycle time, improve artifact quality, increase productivity, and support a broader range of cognitive and creative activities. The reduction in cycle time emerges as the most prominent perceived benefit, reported by more than half of the respondents, reinforcing the view that GenAI tools primarily function as accelerators of SE activities such as coding, debugging, and prototyping. Closely related to cycle time reduction, productivity increase is also reported, though less frequently. This distinction suggests that practitioners may more readily perceive and articulate immediate time savings than higher-level productivity gains, which are more abstract and harder to isolate from contextual factors. When asked specifically (Q17), approximately 95\% of our participants reported a productivity increase. Similarly, Google’s DORA report states that more than 80\% of respondents report a perception that AI has increased their productivity \cite{Google2025b}. Those organizations actively using GenAI in SE have seen an average total productivity improvement of 7–18\% across the software development lifecycle, compared to non-usage of GenAI \cite{Capgemini2024}.

Productivity gains are not even across all SE activities; documentation and coding show the highest time saving followed by debugging and testing \cite{Capgemini2024}. Peng et al. \cite{Peng2023} report that GitHub Copilot significantly impacts productivity, allowing developers to complete a standardized programming task 55.8\% faster than those without it. An evaluation of GitHub Copilot in open-source software found that it positively influences project-level productivity, signaled by increased merged pull requests and improved individual productivity, especially among core developers who derive greater benefits due to higher AI-developer complementarity \cite{Song2024}. While reporting a productivity increase, Peng et al. \cite{Peng2023} and Song et al. \cite{Song2024} did not assess code quality and mentioned this as a limitation to their results. As per the empirical study by Ulfsnes et al. \cite{Ulfsnes2024}, ChatGPT caused spending less time on manual repetitive tasks leading to increased productivity. On the other hand, a study by Becker et al. \cite{Becker2025} found that AI systems can slow down experienced developers by shifting their time from active coding to prompting, waiting on, and reviewing AI outputs, suggesting that AI capabilities in real-world settings may be lower than commonly believed.

Additionally, GenAI can lead to “task-complexity polarization,” a paradox where automation makes easy tasks easier but ironically can make hard tasks even harder, potentially increasing user cognitive workload. \added{Afroz et al. describe a related pattern as “spurious productivity,” in which effort moves from writing code to prompting, debugging, verification, and review \cite{afroz2026fast}. In their study, 84.3\% of frequent users stated that GenAI did not reduce review time. Tabarsi et al. similarly report a productivity-quality paradox: developers completed routine work faster but often discarded generated code and spent additional effort evaluating and integrating it \cite{tabarsi2027llms}.} These findings suggest that productivity gains are dependent on several factors, such as SE activity, task complexity, experience of practitioners on both SE (to validate outputs) and GenAI (to obtain better output), and even programming language for code-related tasks \cite{Peng2023}. \added{Therefore, our findings overlap with evidence of faster task completion but conflict with studies showing that GenAI can redistribute effort rather than remove it.}

Quality improvement is reported as a benefit by more than one-third of our respondents. Additionally, 82 of our participants perceive an improvement in their work quality when GenAI tools are used. On the other hand, our respondents also highlighted several challenges regarding GenAI outputs, i.e., inaccuracy, validation need, low quality outputs due to lack of context understanding, forgetting, and unwanted suggestions. Google’s DORA report states that 59\% of their respondents observed that AI has positively impacted their code quality \cite{Google2025b}. Conversely, 10\% of their respondents perceived any negative impacts on their code quality as a result of AI use \cite{Google2025b}. Despite these perceived quality improvements reported by respondents, several empirical studies report quality problems related to GenAI outputs. An empirical study by Tihanyi et al. \cite{Tihanyi2025} found out that 62\% of LLM-generated code is vulnerable. Liu et al. \cite{Liu2024} reports that code generated by ChatGPT has vulnerabilities. Additionally, ChatGPT’s ability to directly fix erroneous code with a multi-round fixing process to achieve correct functionality is relatively weak \cite{Liu2024}. LLMs also replicate buggy code patterns frequently during code completion \cite{Guo2025}. \added{These conflicting results extend to maintainability. Cursor, a popular LLM agent assistant, adoption produced short-term velocity gains but also persistent technical debt that reduced later development velocity \cite{he2026speed}. AI agents also generated more redundant code than human developers, while reviewers were less critical of AI-generated pull requests \cite{huang2026more}. This suggests that plausible-looking code can hide design and maintenance problems.}

While the surveys, including ours, generally report an improvement in perceived productivity and quality, these results must be interpreted carefully. Especially when we consider limited utilization of objective metrics to measure productivity and quality (as reported in Section \ref{sectionRQ2.4}), these reported perceived improvements become even more questionable. The self-report literature on technology adoption \cite{schaupp2010file, ciriello2023yea} and particularly on AI adoption \cite{yu2026cognitiveoffloadingspeedupillusion} document an optimism bias in perceived gains. \added{Existing studies support both sides of this discussion. Some studies confirm measurable productivity gains \cite{cui2026effects, demirer2026writing, tomaz2026impacts}, while others show perceived speedups alongside actual delays, additional review work, and technical debt \cite{afroz2026fast, Becker2025, he2026speed, huang2026more}. These differences may arise from the task, work setting, measurement level, and time period considered.}

Future research that triangulates these perceptions with repository data, telemetry, or controlled experimental designs would substantially strengthen the evidentiary basis for the productivity and quality claims that currently rest on self-report alone. \added{Such research should also measure review effort, rework, vulnerabilities, technical debt, releases, and long-term maintainability. This would help distinguish immediate task acceleration from sustainable improvements in software delivery.} According to the SEI’s AI Adoption Maturity Model Overview, measurement plays a central role in effective AI adoption: It is essential not only for evaluating current AI adoption maturity, but also for guiding strategic planning, tracking progress, and achieving predictable, sustainable outcomes from AI initiatives \cite{Ozkaya2025}.

The use of GenAI tools in SE introduces significant and complex concerns across privacy, ethics, copyright, and IP. Our results reveal that these concerns are not geography specific. According to our data, three respondents from India, the Netherlands and Türkiye reported that they do not utilize GenAI tools due to such concerns. Four respondents from Brazil, Jordan, Philippines, and Russia see these concerns as factors making GenAI adoption challenging. In addition, 12 more unique participants raised these concerns in response to the question on additional important issues (Q30). These participants were from Australia, Azerbaijan, Brazil, Egypt, Jordan, Malaysia, Philippines, Taiwan, Tanzania, Türkiye, and the US (2). The literature includes several findings on these global concerns. A primary concern is the risk of data leakage and the unauthorized use of proprietary information \cite{Ahmed2025}. Since many GenAI services are cloud-based, sending sensitive data, such as source code or requirements, to third-party providers raises confidentiality risks \cite{Banh2025, Russo2024a}. LLMs may inadvertently memorize and reproduce sensitive or proprietary details present in their training data \cite{Belzner2023, Lo2023}. There is also a legal uncertainty regarding the ownership of content generated by GenAI \cite{Ahmed2025, Banh2025}. LLMs are often trained on vast corpora, including open-source projects under various licenses, leading to the risk that the generated output may contain verbatim or slightly modified copies of copyrighted content \cite{Ahmed2025}. This raises serious concerns about potential copyright infringement and adherence to license requirements \cite{Ahmed2025, Hassan2024a}. Organizations must navigate these challenges via robust governance and mechanisms.

\textit{Institutionalization of tools and techniques.} The rapid evolution of AI capabilities has attracted the attention of policymakers. Key political players, such as the US and EU, have enacted regulations and many governments announced major investments in AI infrastructure. This focus on regulation and investment is an important step towards fulfilling the vital need of realizing the transformative potential of AI as well as managing the associated risks \cite{StanfordHAI2025}. Similarly, at company-level, the integration of GenAI into SE practices needs effective organizational governance and support. As per our findings, nearly two-thirds of our respondents receive organizational support for GenAI tool use. This indicates that GenAI is moving beyond individual experimentation toward being recognized as a strategic enterprise asset.

The two most prevalent forms of support in our study are providing direct access to GenAI tools (81\%) and integration of GenAI tools into existing working environments (51\%). Capgemini Research Institute \cite{Capgemini2024} reported that only 27\% of organizations have the required platforms and tools to utilize GenAI. MIT report “State of GenAI in Business” reveals that 40\% of companies say they purchased an official LLM subscription \cite{Challapally2025}. The lack of formal organizational provision does not necessarily deter employees from utilizing AI tools. The MIT report revealed a “shadow AI economy” in which employees use personal ChatGPT accounts, Claude subscriptions, and other consumer tools to automate significant portions of their jobs, often without IT knowledge or approval \cite{Challapally2025}. Despite the fact that only 40\% of the companies surveyed formally purchased an official LLM subscription, more than 90\% of the employees of those same companies reported regular use of GenAI tools for their work \cite{Challapally2025}. Similarly, Capgemini Research Institute \cite{Capgemini2024} reports that of those software professionals who use GenAI, 63\% use unauthorized tools.

Employees need to learn how to utilize GenAI tools in addition to having access to them. Less than half of our respondents, i.e., 45\%, obtain training provided by their organization. Almost similar to our results, Capgemini Research Institute \cite{Capgemini2024} reported that approximately 40\% of employees receive training from their organizations. Similarly, nearly three in five organizations have no training program for GenAI to upskill and reskill their employees. As a result of this, some software professionals (32\%) are trying to cover the lack of organizational support and getting trained on GenAI independently \cite{Capgemini2024}. Lack of training can have several negative consequences including underuse of powerful GenAI tools \cite{Bain2025} or increase in technical debt \cite{Anderson2025}.

A total of 41\% of our respondents reported that their organization published written policies and guidelines. Very similarly, Capgemini Research Institute \cite{Capgemini2024} found out that governance and policies are in place in 44\% of organizations they surveyed. As per Deloitte’s report on GenAI, 56\% of surveyed organizations say that establishing a governance framework for the use of GenAI is key to mitigating risk related to trust in the tool \cite{Muratovic2024}. Establishing acceptable-use policies for AI is strongly associated with a substantial increase in adoption; Google’s DORA survey \cite{Google2025a} indicates a 451\% increase in AI adoption in organizations that create these policies compared to those that do not. Implementing clear organizational policies regarding AI usage offers developers a structured framework necessary for confident, responsible, and effective tool utilization. This clarity is vital for establishing the psychological safety required for effective experimentation, which consequently reduces friction and amplifies AI’s positive impact on both individual effectiveness and overall organizational performance \cite{Google2025b}. While such policies and guidelines are critical for responsible adoption of GenAI, more detailed and actionable guidelines are also needed. For instance, as AI-assisted implementation tools are becoming more common, guidelines are needed to tackle technical debt in AI-generated code, especially for novice developers \cite{Anderson2025}.

To complement central governance, companies can involve employees in integrating AI capabilities into company processes. Pinterest, a social media company, runs hackathons to promote collaboration and speed up software development. Just a few months after the release of ChatGPT, they created an internal chatbot as a result of a mini hackathon and made it available to its employees for wider use. Keeping employees closely involved in the development of internal AI tools not only facilitates employee buy-in, but also fuels ground-up innovation \cite{Chandonnet2025}.

Both our data and other reports, such as DefineX’s “Beyond Code” report \cite{DefineX2025}, show that even though some organizations have introduced basic enablers, there is still much more room to improve institutionalization of tools and techniques for GenAI. Institutionalization requires an organizational-specific systematic framework. This framework can be grounded on existing standards such as ISO/IEC 42001:2023 (AI Management System) \cite{ISO42001} and ISO/IEC 38507:2022 (Governance implications of AI use by organizations) \cite{ISO38507}, as well as AI adoption maturity models like \cite{Ozkaya2025}. The framework must include a structured roadmap that is distributed throughout the organization in the form of an AI strategy \cite{Ozkaya2025}. The competency development pillar \cite{ISO42001, Ozkaya2025} addresses the human capital dimension of GenAI adoption, covering structured training, prompt engineering as a formal SE competency, and role-specific upskilling that differentiates, for example, the needs of developers, testers, requirement engineers, and architects. Workflow re-engineering \cite{Ozkaya2025} and the provision of organizational access to GenAI tools are critical to the readiness of processes and technological infrastructure for GenAI adoption. Not only providing access to GenAI tools, but also the compatibility of AI tools within SE workflows predominantly drives their adoption \cite{Russo2024a}. The governance and policy pillar establishes the organizational rules of engagement for GenAI use, including acceptable-use policies, security and data privacy guidelines that address the risk of inadvertent disclosure of proprietary code or sensitive data, copyright and IP frameworks, and ethical guidelines covering concerns such as overreliance and transparency \cite{ISO42001, ISO38507}. The measurement and accountability pillar, related to the performance evaluation clause in \cite{ISO42001}, covers the definition and tracking of objective metrics for productivity, quality, and SE artifact size in GenAI-assisted workflows; the establishment of baselines against which AI-induced changes can be measured; and the use of KPIs tied to GenAI adoption objectives. The continuous improvement and adaptation pillar, related to the improvement clause in \cite{ISO42001}, ensures that the GenAI adoption practices are regularly reviewed and updated in response to the evolving capabilities of the tools, emerging risks, and measured results. It includes mechanisms for capturing practitioner feedback on tool effectiveness, processes for updating policies and training as the GenAI landscape evolves, and organizational structures (such as dedicated GenAI experts or center of excellence) that provide sustained institutional knowledge. We emphasize that this framework is preliminary and empirically motivated rather than prescriptively validated.

\textit{Potential impact on the SE community.} The responses related to RQ4 capture practitioner expectations about futures that have not yet materialized and that are subject to considerable uncertainty. The pace of GenAI development, the plasticity of labor markets, and the complexity of organizational change make five-year predictions inherently unstable. We therefore discuss these findings as attitudinal signals rather than as predictive evidence about what the SE profession will look like in 2030.

GenAI tools exhibit great potential for automating knowledge work in various sectors. Previous projections estimated that technical automation could affect approximately 50\% of employee time; this figure has accelerated to a range of 60\% to 70\% due to GenAI’s enhanced natural-language processing capabilities \cite{McKinsey2023}.

\replaced{}{Most of our respondents (79\%) expect or anticipate that GenAI will redefine their current role, rather than replace it.} Capgemini Research Institute \cite{Capgemini2024} reports that 59\% of software professionals do not see GenAI as a threat to job displacement, while 22\% stay neutral. 19\% think their jobs are under threat, which is very close to our finding, i.e., 21\%. While two-thirds of developers (64\%) believe that AI is not a threat to their job, there is a slight decrease in confidence versus 2024, i.e., 68\% \cite{StackOverflow2025}. \added{Most of our respondents (79\%) expect GenAI to redefine their current role rather than replace it. This finding is consistent with recent research showing that GenAI is changing role boundaries and shifting work toward more strategic, creative, and complex tasks. Product managers are increasingly performing technical activities previously handled by software engineers, such as prototyping and querying \cite{ulloa2026product}, while product owners, developers, and Scrum masters are evolving into more strategic, innovative, and supportive roles \cite{diebold2025backlogs}.}

Despite the high confidence in individual job security, 54\% of our participants anticipate the overall SE job market will shrink within five years, likely due to efficiency gains. MIT’s “State of GenAI in Business” report cites that more than 80\% of executives anticipate reduced hiring volumes within 24 months in the technology sector due to GenAI impact \cite{Challapally2025}. In contrast with these expectations, the Future of Jobs Survey by World Economic Forum \cite{WEF2025} reported that the fastest growing jobs by 2030 include roles such as Big Data Specialist, FinTech Engineers, AI and Machine Learning Specialists and Software and Applications Developers, projected by surveyed employers. The expected net growth for these roles is over 50\%. For instance, companies in Germany and Canada expect their business models to be reshaped by digital technologies including AI. To prepare for these changes, companies plan to actively hire staff for roles like software developers, UI/UX designers, AI and ML specialists, and security management specialists.

Our findings highlight an awareness of the need for upskilling, coupled with high confidence in the ability to adapt. An overwhelming 84\% of our participants are confident they can acquire the necessary skills to utilize GenAI tools effectively. This readiness aligns with Gartner’s estimation stating that 80\% of the SE workforce is projected to require specialized GenAI-related upskilling through 2027 \cite{Gartner2024}.

Our survey found that 45\% of respondents do not expect negative impacts on compensation, yet a significant minority (31\%) express concern that efficiency gains could lead to downward pressure on salaries. The automation of repetitive software engineering tasks through artificial intelligence is expected to reshape the SE profession by displacing or transforming roles centered on routine activities \cite{Tschang2021}. Tasks such as manual testing and basic coding are increasingly susceptible to automation, leading to the commoditization of traditional coding skills and potential downward pressure on wages or the elimination of certain positions \cite{Ahmed2025, Ebert2023}. Conversely, demand is projected to grow for professionals with strong architectural, design, and problem-solving capabilities, as these higher-order skills enable the creation of innovative solutions and strategic oversight that cannot be readily automated \cite{Ahmed2025}.

The effect of GenAI on social dynamics in the workplace is the least consensus-driven area of our study, with responses closely split (41\% agreeing and 37\% disagreeing) on whether GenAI may result in less social interactions. While most Europeans (60\%+) view robots and AI positively due to their potential to alleviate repetitive tasks and enhance decision accuracy, 84\% think that their adoption needs careful management \cite{EuropeanCommission2024}. Notably, although some professional benefits are recognized, respondents are wary of their impact on social areas like communication among colleagues. This highlights an area of ongoing debate, where the convenience of consulting a tool may reduce human interaction, even as the new roles created by GenAI may necessitate more collaboration on complex problems. If GenAI significantly reduces human interaction, it may lead to negative results. If team members reduce their interactions in favor of focusing on individual tasks, they may be isolated from the team focusing on their own goals and plans rather than team goals \cite{Ulfsnes2024}. In addition, people may be less satisfied with their work, as “helping a co-worker” is generally considered positive and rewarding \cite{Meyer2019}.

\section{Threats to Validity} \label{threatstovalidity}
\textit{Face and Content Validity.} Face and content validity threats include bad instrumentation and inadequate explanation of constructs. To mitigate these threats, we involved five SE professionals in reviewing and evaluating the questionnaire with respect to the format and formulation of the questions. Additionally, one social scientist who is expert on survey research provided feedback on face validity.

\textit{Criterion Validity.} We clearly and concisely explained the purpose and content of the questionnaire on the introduction page. We explicitly asked potential respondents whether they are willing to fill out the questionnaire. We considered only answers meeting the quality criterion, i.e., providing one SE activity for which a GenAI tool is used for.

\textit{Construct Validity.} \replaced{The dataset analyzed in this paper was collected via a questionnaire, and construct validity threats are inherent to any self-reported, questionnaire-based study. One such threat is the risk of misunderstood questions, which could lead to incomplete or incorrect responses. Related to this is the risk that respondents interpret and answer through their own subjective lenses; for instance, a respondent might perceive changes to their own role differently than changes to other roles. Beyond question clarity, self-reported data is also subject to recall bias, where respondents may inaccurately remember past events or experiences \cite{coughlin1990recall}, and to social desirability bias \cite{krumpal2013determinants} and self-assessment bias \cite{mabe1982validity}, where respondents may over- or under-report certain behaviors, attitudes, or competencies, whether consciously or not. To mitigate these threats to the extent possible, we conducted a pilot study with five experienced SE practitioners, paying special attention to reducing question-level bias and potential misunderstandings. Additionally, we involved a social scientist who is an expert on survey research to verify the questions. We note, of course, that these steps primarily address question clarity and design, and do not fully eliminate threats such as recall bias, social desirability bias, and self-assessment bias, which are inherent to self-reported data and should be considered when interpreting our results.}{The dataset analyzed in this paper was collected via a questionnaire. One of the main threats of survey-based research is the risk of misunderstood questions leading to incomplete or wrong responses. Related to that threat is also that the respondents answer through their very own lenses and might well perceive, for instance, changes on their own role differently than on other roles. To mitigate this risk, a pilot study with five experienced SE practitioners was conducted, paying special attention to reduce the level of bias and potential misunderstandings as much as possible. Additionally, a social scientist who is an expert on survey research verified the questions.}

\textit{External Validity.} We used convenience and snowball sampling strategies to identify potential respondents. \replaced{To decrease the risk of the dataset being dominated by a small number of countries}{To decrease the risk of having too many data points from some countries}, we also employed purposive sampling. We continuously tracked the number of responses per country and stopped promoting the survey in a country \replaced{once we had obtained a reasonable number of responses relative to other countries}{when we got enough responses}. For instance, we stopped data collection in Brazil and Türkiye when we had approximately 20 responses, which is slightly below the number of responses obtained from the US. By doing so, \replaced{we aimed to increase the geographic diversity of our sample and avoid a dataset dominated by one or two large, easily-reached populations, rather than to achieve a sample that is proportionally representative of the global SE workforce}{we aimed at obtaining a dataset that is geographically as unbiased as possible}. Furthermore, we did a quality check on the responses as explained in Section \ref{dataanalysis}. 

Although respondents were drawn from 38 countries, the regional breakdown provided in Table \ref{tab:responses_by_regions} confirms that the sample is geographically diverse, however, it is not regionally balanced. In particular, technology hub countries within North America and parts of Asia may be under- or overrepresented relative to their share of the global SE workforce.
As a result, the findings may not be generalized to a broader SE population. Although non-probability sampling, including convenience, purpose, and snowball, is used frequently in SE research, it introduces a threat to generalizability, subjectivity, and can lead to sampling a small, highly-interconnected subset of a larger SE population \cite{Baltes2022}.

\textit{Reliability.} \replaced{To account for the uncertainty introduced by our sample size and to report our estimates conservatively, we used bootstrapping and report the resulting confidence intervals. It is important to note that bootstrapping does not address limitations arising from non-probability sampling; it only estimates uncertainty conditional on the observed sample, and does not, by itself, support claims of population-level generalizability. Since we bootstrap by resampling from the same non-probability sample, the resulting confidence intervals represent within-sample variability, not population-level variability.}{To deal with the random sampling limitation, we used bootstrapping and conservatively reported confidence intervals. Based on the inferential statistics and the small obtained confidence intervals, we believe that our results are mainly generalizable. It is important to note that the confidence intervals represent within-sample variability, not population-level variability, since we bootstrap with resampling from the same non-probability sample.} Therefore, we avoided further generalizability claims throughout the paper due to the aforementioned limitations. Replications should be conducted to further strengthen the statistical generalizability.

Another reliability aspect concerns \replaced{coding}{inter-observer} reliability, which we improved by including two independent peer reviews in all our qualitative analysis procedures and making all the data and analyses openly available online.

\section{Conclusions} \label{conclusions}
This study provides a comprehensive and empirically grounded overview of the state of GenAI adoption in software engineering, encompassing the status of GenAI adoption, associated benefits and challenges, institutionalization of tools and techniques, and anticipated long-term impacts on industrial SE practice. The findings from our sample show that GenAI tools are commonly and frequently used for SE activities, particularly implementation, verification \& validation, personal assistance, and maintenance. Practitioners generally perceive GenAI as an enabler of cycle time reduction, quality improvement, enhanced support in knowledge work, and productivity gains. These perceptions are consistent and widespread across our geographically diverse sample, lending them credibility as indicators of practitioner experience. However, given that the majority of organizations lack objective productivity and quality metrics, these findings represent self-reported sentiment rather than independently verified performance improvements, and should be interpreted with corresponding caution.

The results also reveal important limitations and risks. Despite the high perceived productivity and quality improvements, objective measurement of GenAI’s impact seems very limited. The challenges, such as incorrect and unreliable outputs, prompt engineering difficulties, output validation overhead, security and privacy concerns, and risk of overreliance, highlight important research directions to close the gap between current GenAI capabilities and the needs of the SE profession. While organizations have taken some actions to support GenAI adoption, there is room for improvement to enhance the benefits as well as managing associated risks. Practitioners mostly expect GenAI to refine their roles rather than replace them, and express concerns about possible skill transformations and long-term job market impacts. These expectations are attitudinal in nature and should be read as signals of how SE practitioners in our sample currently perceive an uncertain future, rather than as evidence-based predictions.

\subsection{Implications for Research}
\added{The findings point challenges concerning how SE should conceptualize work in which GenAI is routinely involved. In particular, the results raise questions about the unit of analysis, established notions of productivity and quality, assurance, and professional expertise.}

\added{\textit{Human-GenAI systems as a unit of analysis.} Our findings show that GenAI is becoming integrated into routine SE work, while practitioners remain responsible for prompting, contextualizing, validating, and correcting its outputs. This challenges perspectives that treat engineers and their tools as clearly separable units. An important research challenge is therefore to conceptualize the human-GenAI system as a unit of analysis and explain how tasks, capabilities, responsibilities, and outcomes are distributed between practitioners and GenAI tools. Mixed-method and multi-source studies combining surveys, industrial observations, repository data, and workflow telemetry can provide evidence for developing and evaluating such models.}

\added{\textit{Reconceptualizing productivity and quality.} The coexistence of strong perceived productivity and quality improvements, validation overhead, and limited objective measurement raises a fundamental question about what constitutes productivity and quality in GenAI-assisted SE. Faster artifact generation may redistribute effort toward prompting, review, correction, integration, or downstream maintenance rather than eliminate it. SE research therefore needs measurement models that account for the complete human-GenAI workflow and distinguish local task acceleration from system-level improvements in engineering outcomes.}

\added{\textit{Assurance for human-GenAI work.} The prevalence of unreliable outputs, validation effort, security concerns, and risks of overreliance creates an assurance problem that extends beyond improving model accuracy. When probabilistic GenAI systems routinely participate in producing software artifacts, assumptions about predictability, trust, and responsibility change. SE research needs to reconsider how verification, trust, accountability, and acceptable evidence should be conceptualized and operationalized for artifacts produced through human-GenAI interaction.}

\added{\textit{Redefining SE expertise.} Finally, practitioners largely expect GenAI to redefine rather than replace their roles, while concerns about overreliance and skill atrophy remain. This raises a broader question about what constitutes SE expertise when parts of artifact production and problem solving can be delegated to GenAI. Longitudinal research should examine how expertise is redistributed among problem framing, artifact creation, delegation, validation, and contextual judgment, and how this redistribution affects learning and professional development over time.}

\replaced{}{\textit{Triangulation through mixed-method and multi-source research.} The findings of this study are grounded in self-reported perceptions and narratives rather than observed practice. Future research should therefore pursue mixed-method designs that combine large-scale surveys with complementary empirical sources: industrial case studies and semi-structured interviews can probe the organizational and contextual factors that surveys cannot capture, while repository mining and workflow telemetry can provide objective behavioral evidence against which self-reported productivity and quality claims can be benchmarked. Such multi-source designs would not only validate and deepen the findings reported here, but would also help distinguish between what practitioners believe GenAI is doing to their work and what it is demonstrably doing.}

\replaced{}{\textit{Structured approach to measurement of GenAI impact.} One of the most obvious research gaps revealed by this study is the absence of structured approaches to measurement of productivity and quality in GenAI-assisted SE. Over 58\% of practitioners report using no objective metric and those who rely predominantly on agile metrics such as story points and velocity, which were not designed to capture AI-induced changes. Researchers should prioritize the development and validation of measurement frameworks tailored to GenAI-assisted SE considering various dimensions of productivity \mbox{\cite{sikand2024much, Coutinho2024}}.}

\replaced{}{\textit{Institutionalization and governance effectiveness.} Organizational governance is vital to harness the power of GenAI while managing its risks \cite{schneider2024governance}. It is important to understand the impact of governance mechanisms on outcome improvement and risk management. Therefore, longitudinal and comparative studies that link organizational governance maturity to measurable SE outcomes would make a significant contribution.}

\replaced{}{\textit{Long-term skill and workforce effects.} Concerns about overreliance, skill atrophy, and job market contraction are present in our data, but remain speculative. Longitudinal studies tracking skill development and workforce composition over time—particularly for early-career practitioners who are most vulnerable to cognitive offloading—are needed to move this debate from prediction to evidence. Not only companies but also educational institutions must adapt their curricula based on such longitudinal studies when needed \cite{mastropaolo2024rise}.}

\subsection{Implications for Practice}
\textit{For individual practitioners.} GenAI tools are widely adopted in SE with approximately four out of five respondents reporting active use. Of active users, 65\% are using GenAI daily, suggesting that GenAI is moving from an optional tool to a professional expectation in many SE roles. Practitioners who have not yet adopted GenAI tools should be aware that the primary barriers—skill and time constraints (23\%)—are addressable through low-cost self-directed learning. At the same time, users should pay attention to the risks of overreliance: the same tools that reduce cycle time can erode critical thinking and debugging skills, particularly among less experienced developers. Developing validation habits and maintaining awareness of when GenAI output requires closer examination is becoming a core professional competency.

\textit{For team leads and engineering managers.} The measurement gap identified in our study has direct managerial consequences. Without objective baselines and metrics, teams cannot distinguish real productivity gains from perceived ones, cannot identify where GenAI is introducing technical debt, and cannot make evidence-based decisions about tool investment. We recommend that teams establish lightweight measurement practices—even simple ones such as tracking defect rates, code review turnaround times, or rework frequency—before and after GenAI adoption, to enable before-after comparisons. Furthermore, given that prompt engineering proficiency heavily determines output quality, teams should consider structured peer-learning arrangements or internal prompt libraries to accelerate skill development.

\textit{For organizational leaders.} Our data reveal a clear institutionalization gap: organizations are much better at providing access to GenAI tools than at governing their use. Only 41\% have published policies and guidelines, only 45\% provide training, and fewer than one in five set and track KPIs tied to GenAI usage. Evidence from Google's DORA survey \cite{Google2025a} indicates that organizations with clear acceptable-use policies see dramatically higher adoption rates and better outcomes. We recommend that organizations treat GenAI governance as a strategic priority.

\textit{For policymakers and professional bodies.} The privacy, security, copyright, and IP concerns surfaced in our data were reported by practitioners across all continents. This points to the need for internationally coordinated regulatory frameworks that give organizations clear guidance on responsible GenAI use without blocking adoption. Professional bodies in SE should also consider updating competency frameworks and certification standards to include GenAI literacy and prompt engineering as recognized skills.

\section*{Declaration of Generative AI and AI-assisted technologies in the writing process}
We declare that AI solutions (i.e., Gemini, ChatGPT, Claude, and Gemini Notebook) were used in the writing process. Gemini was used for text summarization, rephrasing, grammar checking, and producing Python scripts to be used for data analysis. ChatGPT and Claude were used for text summarization, rephrasing, and grammar checking. Gemini Notebook was used to extract and summarize information from papers. We thus declare that we (the authors) have fully authored the content of this article and take full responsibility for the content of the publication.

\section*{Data Availability}
The questionnaire, the collected data, and the quantitative and qualitative data analysis artifacts, including Python scripts that produce all results from the collected data, are available in our online open science repository \cite{Giray2026}.

\section*{Acknowledgments}
We would like to thank all participants who responded to our questionnaire and made this study possible.

\bibliographystyle{ACM-Reference-Format}
\bibliography{GenAIforSE-References}

@inproceedings{afroz2026fast,
  title={The fast and spurious: Developer productivity with genai},
  author={Afroz, Sadia and Feng, Zixuan and Menezes, Tyler and Kimura, Katie and Trinkenreich, Bianca and Steinmacher, Igor and Sarma, Anita},
  booktitle={Proceedings of the 34th ACM International Conference on the Foundations of Software Engineering},
  pages={1832--1841},
  year={2026}
}

@article{Ahmed2025,
  author = {Ahmed, I. and Aleti, A. and Cai, H. and Chatzigeorgiou, A. and He, P. and Hu, X. and others},
  title = {Artificial Intelligence for Software Engineering: The Journey so far and the Road ahead},
  journal = {ACM Transactions on Software Engineering and Methodology},
  volume = {34},
  number = {5},
  pages = {1--27},
  year = {2025}
}

@article{Akdur2018,
  title={A survey on modeling and model-driven engineering practices in the embedded software industry},
  author={Akdur, Deniz and Garousi, Vahid and Demir{\"o}rs, Onur},
  journal={Journal of Systems Architecture},
  volume={91},
  pages={62--82},
  year={2018},
  publisher={Elsevier}
}

@article{Anderson2025,
  author = {Anderson, E. and Parker, G. and Tan, B.},
  title = {The Hidden Costs of Coding With Generative AI},
  journal = {MIT Sloan Management Review},
  volume = {67},
  number = {1},
  pages = {12--14},
  year = {2025}
}

@article{Baltes2022,
  author = {Baltes, S. and Ralph, P.},
  title = {Sampling in software engineering research: A critical review and guidelines},
  journal = {Empirical Software Engineering},
  volume = {27},
  number = {4},
  pages = {94},
  year = {2022}
}

@article{Banh2025,
  author = {Banh, L. and Holldack, F. and Strobel, G.},
  title = {Copiloting the Future: How Generative AI Transforms Software Engineering},
  journal = {Information and Software Technology},
  pages = {107751},
  year = {2025}
}

@inproceedings{Belzner2023,
  author = {Belzner, L. and Gabor, T. and Wirsing, M.},
  title = {Large language model assisted software engineering: prospects, challenges, and a case study},
  booktitle = {International Conference on Bridging the Gap between AI and Reality},
  pages = {355--374},
  year = {2023},
  publisher = {Springer Nature Switzerland}
}

@inproceedings{Coutinho2024,
  author = {Coutinho, M. and Marques, L. and Santos, A. and Dahia, M. and França, C. and de Souza Santos, R.},
  title = {The role of generative ai in software development productivity: A pilot case study},
  booktitle = {Proceedings of the 1st ACM International Conference on AI-Powered Software},
  pages = {131--138},
  year = {2024}
}

@article{Damyanov2024,
  author = {Damyanov, I. and Tsankov, N. and Nedyalkov, I.},
  title = {Applications of Generative Artificial Intelligence in the Software Industry},
  journal = {TEM Journal},
  volume = {13},
  number = {4},
  year = {2024}
}

@article{Ebert2023,
  author = {Ebert, C. and Louridas, P.},
  title = {Generative AI for software practitioners},
  journal = {IEEE Software},
  volume = {40},
  number = {4},
  pages = {30--38},
  year = {2023}
}

@book{Efron1994,
  author = {Efron, B. and Tibshirani, R. J.},
  title = {An introduction to the bootstrap},
  publisher = {Chapman and Hall/CRC},
  year = {1994}
}

@inproceedings{felder2026adoption,
  title={Adoption of Generative Artificial Intelligence in the German Software Engineering Industry: An Empirical Study},
  author={Felder, Ludwig and Eisenreich, Tobias and Fischer, Mahsa and Wagner, Stefan and Chen, Chunyang},
  booktitle={Proceedings of the 34th ACM International Conference on the Foundations of Software Engineering},
  pages={334--344},
  year={2026}
}

@article{Garousi2015,
  title={A survey of software engineering practices in Turkey},
  author={Garousi, Vahid and Co{\c{s}}kun{\c{c}}ay, Ahmet and Betin-Can, Aysu and Demir{\"o}rs, Onur},
  journal={Journal of Systems and Software},
  volume={108},
  pages={148--177},
  year={2015},
  publisher={Elsevier}
}

@article{Garousi2019a,
  author = {Garousi, V. and Giray, G. and Tuzun, E.},
  title = {Understanding the knowledge gaps of software engineers: An empirical analysis based on SWEBOK},
  journal = {ACM Transactions on Computing Education (TOCE)},
  volume = {20},
  number = {1},
  pages = {1--33},
  year = {2019}
}

@inproceedings{Geng2024,
  author = {Geng, M. and Wang, S. and Dong, D. and Wang, H. and Li, G. and Jin, Z. and others},
  title = {Large language models are few-shot summarizers: Multi-intent comment generation via in-context learning},
  booktitle = {Proceedings of the 46th IEEE/ACM International Conference on Software Engineering},
  pages = {1--13},
  year = {2024}
}

@article{Giray2021a,
  author = {Giray, G.},
  title = {A software engineering perspective on engineering machine learning systems: State of the art and challenges},
  journal = {Journal of Systems and Software},
  volume = {180},
  pages = {111031},
  year = {2021}
}

@article{Giray2021b,
  author = {Giray, G.},
  title = {An assessment of student satisfaction with e-learning: An empirical study with computer and software engineering undergraduate students in Turkey under pandemic conditions},
  journal = {Education and Information Technologies},
  volume = {26},
  number = {6},
  pages = {6651--6673},
  year = {2021}
}

@dataset{Giray2026,
  author       = {Giray, Görkem and
                  Demirors, Onur and
                  Kalinowski, Marcos and
                  Méndez Fernández, Daniel},
  title        = {An Empirical Study of Generative AI Adoption in
                   Software Engineering
                  },
  month        = oct,
  year         = 2026,
  publisher    = {Zenodo},
  version      = 2,
  doi          = {10.5281/zenodo.23136594},
  url          = {https://doi.org/10.5281/zenodo.23136594},
}

@phdthesis{Glushkova2023,
  author = {Glushkova, D.},
  title = {The influence of Artificial intelligence on productivity in Software development},
  school = {Politecnico di Torino},
  year = {2023}
}

@inproceedings{Hassan2024a,
  author = {Hassan, A. E. and Lin, D. and Rajbahadur, G. K. and Gallaba, K. and Cogo, F. R. and Chen, B. and others},
  title = {Rethinking software engineering in the era of foundation models: A curated catalogue of challenges in the development of trustworthy fmware},
  booktitle = {Companion Proceedings of the 32nd ACM International Conference on the Foundations of Software Engineering},
  pages = {294--305},
  year = {2024}
}

@mastersthesis{Hassan2024b,
  author = {Hassan, M. A.},
  title = {Impact of adopting AI tools by software developers towards productivity and sustainability},
  school = {Lappeenranta–Lahti University of Technology LUT},
  year = {2024}
}

@article{Hou2024,
  author = {Hou, X. and Zhao, Y. and Liu, Y. and Yang, Z. and Wang, K. and Li, L. and others},
  title = {Large language models for software engineering: A systematic literature review},
  journal = {ACM Transactions on Software Engineering and Methodology},
  volume = {33},
  number = {8},
  pages = {1--79},
  year = {2024}
}

@misc{ISO25059,
  title = {{ISO/IEC} International Standard - Software engineering - Systems and software Quality Requirements and Evaluation ({SQuaRE}) - Quality model for {AI} systems, {ISO/IEC} 25059:2023},
  year = {2023}
}

@misc{ISO12207,
  title = {{ISO/IEC/IEEE} International Standard - Systems and software engineering - Software life cycle processes, {ISO/IEC/IEEE} 12207:2017(E)},
  year = {2017}
}

@misc{ISO42001,
  title = {{ISO/IEC} International Standard - Information technology — Artificial intelligence — Management system, {ISO/IEC} 42001:2023},
  year = {2023}
}

@misc{ISO38507,
  title = {{ISO/IEC} International Standard - Information technology — Governance of IT — Governance implications of the use of artificial intelligence by organizations, {ISO/IEC} 38507:2022},
  year = {2022}
}

@inproceedings{Jahic2024,
  author = {Jahić, J. and Sami, A.},
  title = {State of Practice: LLMs in Software Engineering and Software Architecture},
  booktitle = {2024 IEEE 21st International Conference on Software Architecture Companion (ICSA-C)},
  pages = {311--318},
  year = {2024},
  publisher = {IEEE}
}

@article{Kalinowski2025,
  author = {Kalinowski, M. and Mendez, D. and Giray, G. and Alves, A. P. S. and Azevedo, K. and Escovedo, T. and others},
  title = {Naming the pain in machine learning-enabled systems engineering},
  journal = {Information and Software Technology},
  pages = {107866},
  year = {2025}
}

@article{Khemka2024,
  author = {Khemka, M. and Houck, B.},
  title = {Toward Effective AI Support for Developers: A survey of desires and concerns},
  journal = {Communications of the ACM},
  volume = {67},
  number = {11},
  pages = {42--49},
  year = {2024}
}

@inproceedings{Krishna2024,
  author = {Krishna, M. and Gaur, B. and Verma, A. and Jalote, P.},
  title = {Using LLMs in software requirements specifications: an empirical evaluation},
  booktitle = {2024 IEEE 32nd International Requirements Engineering Conference (RE)},
  pages = {475--483},
  year = {2024},
  publisher = {IEEE}
}

@article{Kuhail2024,
  author = {Kuhail, M. A. and Mathew, S. S. and Khalil, A. and Berengueres, J. and Shah, S. J. H.},
  title = {“Will I be replaced?” Assessing ChatGPT's effect on software development and programmer perceptions of AI tools},
  journal = {Science of Computer Programming},
  volume = {235},
  pages = {103111},
  year = {2024}
}

@article{Lei2003,
  author = {Lei, S. and Smith, M. R.},
  title = {Evaluation of several nonparametric bootstrap methods to estimate confidence intervals for software metrics},
  journal = {IEEE Transactions on Software Engineering},
  volume = {29},
  number = {11},
  pages = {996--1004},
  year = {2003}
}

@inproceedings{Li2024,
  author = {Li, M. M. and Dickhaut, E. and Bruhin, O. and Wache, H. and Weritz, P.},
  title = {More than just efficiency: Impact of generative AI on Developer productivity},
  booktitle = {Americas Conference on Information Systems (AMCIS)},
  year = {2024}
}

@inproceedings{Liang2024,
  author = {Liang, J. T. and Yang, C. and Myers, B. A.},
  title = {A large-scale survey on the usability of ai programming assistants: Successes and challenges},
  booktitle = {Proceedings of the 46th IEEE/ACM international conference on software engineering},
  pages = {1--13},
  year = {2024}
}

@article{Liu2024,
  author = {Liu, Z. and Tang, Y. and Luo, X. and Zhou, Y. and Zhang, L. F.},
  title = {No need to lift a finger anymore? assessing the quality of code generation by chatgpt},
  journal = {IEEE Transactions on Software Engineering},
  year = {2024}
}

@inproceedings{Lo2023,
  author = {Lo, D.},
  title = {Trustworthy and synergistic artificial intelligence for software engineering: Vision and roadmaps},
  booktitle = {2023 IEEE/ACM International Conference on Software Engineering: Future of Software Engineering (ICSE-FoSE)},
  pages = {69--85},
  year = {2023},
  publisher = {IEEE}
}

@article{Lunneborg2001,
  author = {Lunneborg, C. E.},
  title = {Bootstrap inference for local populations},
  journal = {Drug information journal},
  volume = {35},
  number = {4},
  pages = {1327--1342},
  year = {2001}
}

@article{Martinovic2025,
  author = {Martinović, B. and Rozić, R.},
  title = {Perceived Impact of AI-Based Tooling on Software Development Code Quality},
  journal = {SN Computer Science},
  volume = {6},
  number = {1},
  pages = {63},
  year = {2025}
}

@article{Meyer2019,
  author = {Meyer, A. N. and Barr, E. T. and Bird, C. and Zimmermann, T.},
  title = {Today was a good day: The daily life of software developers},
  journal = {IEEE Transactions on Software Engineering},
  volume = {47},
  number = {5},
  pages = {863--880},
  year = {2019}
}

@article{Molleri2020,
  author = {Molléri, J. S. and Petersen, K. and Mendes, E.},
  title = {An empirically evaluated checklist for surveys in software engineering},
  journal = {Information and Software Technology},
  volume = {119},
  pages = {106240},
  year = {2020}
}

@article{Ozkaya2023,
  author = {Ozkaya, I.},
  title = {Application of large language models to software engineering tasks: Opportunities, risks, and implications},
  journal = {IEEE Software},
  volume = {40},
  number = {3},
  pages = {4--8},
  year = {2023}
}

@article{Partridge1988,
title = {Artificial intelligence and software engineering: a survey of possibilities},
journal = {Information and Software Technology},
volume = {30},
number = {3},
pages = {146-152},
year = {1988},
note = {The Software Life Cycle},
issn = {0950-5849},
doi = {https://doi.org/10.1016/0950-5849(88)90061-4},
url = {https://www.sciencedirect.com/science/article/pii/0950584988900614},
author = {Derek Partridge},
}

@article{Pezze2025,
  author = {Pezzè, M. and Abrahão, S. and Penzenstadler, B. and Poshyvanyk, D. and Roychoudhury, A. and Yue, T.},
  title = {A 2030 Roadmap for Software Engineering},
  journal = {ACM Transactions on Software Engineering and Methodology},
  volume = {34},
  number = {5},
  pages = {1--55},
  year = {2025}
}

@article{Ralph2020,
  author = {Ralph, P. and Baltes, S. and Adisaputri, G. and Torkar, R. and Kovalenko, V. and Kalinowski, M. and others},
  title = {Pandemic programming: How COVID-19 affects software developers and how their organizations can help},
  journal = {Empirical Software Engineering},
  volume = {25},
  pages = {4927--4961},
  year = {2020}
}

@article{Rech2004,
  author = {Rech, J. and Althoff, K. D.},
  title = {Artificial intelligence and software engineering: Status and future trends},
  journal = {KI},
  volume = {18},
  number = {3},
  pages = {5--11},
  year = {2004}
}

@book{Russell2013,
  author = {Russell, S. and Norvig, P.},
  title = {Artificial intelligence: a modern approach},
  publisher = {Pearson Education Limited},
  year = {2013}
}

@article{Russo2024a,
  author = {Russo, D.},
  title = {Navigating the complexity of generative ai adoption in software engineering},
  journal = {ACM Transactions on Software Engineering and Methodology},
  volume = {33},
  number = {5},
  pages = {1--50},
  year = {2024}
}

@article{Sergeyuk2025,
  author = {Sergeyuk, A. and Golubev, Y. and Bryksin, T. and Ahmed, I.},
  title = {Using AI-based coding assistants in practice: State of affairs, perceptions, and ways forward},
  journal = {Information and Software Technology},
  volume = {178},
  pages = {107610},
  year = {2025}
}

@inproceedings{Stol2016,
  author = {Stol, K. J. and Ralph, P. and Fitzgerald, B.},
  title = {Grounded theory in software engineering research: a critical review and guidelines},
  booktitle = {Proceedings of the 38th International conference on software engineering},
  pages = {120--131},
  year = {2016}
}

@article{Tenekeci2026,
  title={Automating software size measurement from python code using language models},
  author={Tenekeci, Samet and {\"U}nl{\"u}, H{\"u}seyin and G{\"u}l, Bedir Arda and Kele{\c{s}}, Damla and K{\"u}{\"u}k, Murat and Demir{\"o}rs, Onur},
  journal={Automated Software Engineering},
  volume={33},
  number={1},
  pages={19},
  year={2026},
  publisher={Springer}
}

@article{Tihanyi2025,
  author = {Tihanyi, N. and Bisztray, T. and Ferrag, M. A. and Jain, R. and Cordeiro, L. C.},
  title = {How secure is AI-generated code: a large-scale comparison of large language models},
  journal = {Empirical Software Engineering},
  volume = {30},
  number = {2},
  pages = {1--42},
  year = {2025}
}

@article{Tschang2021,
  author = {Tschang, F. T. and Almirall, E.},
  title = {Artificial intelligence as augmenting automation: Implications for employment},
  journal = {Academy of Management Perspectives},
  volume = {35},
  number = {4},
  pages = {642--659},
  year = {2021}
}

@inproceedings{Ulfsnes2024,
  author = {Ulfsnes, R. and Moe, N. B. and Stray, V. and Skarpen, M.},
  title = {Transforming software development with generative AI: empirical insights on collaboration and workflow},
  booktitle = {Generative AI for effective software development},
  pages = {219--234},
  year = {2024},
  publisher = {Springer Nature Switzerland}
}

@article{Unlu2026,
  title = {Automating software size measurement with language models: Insights from industrial case studies},
  journal = {Journal of Systems and Software},
  volume = {231},
  pages = {112638},
  year = {2026},
  issn = {0164-1212},
  doi = {https://doi.org/10.1016/j.jss.2025.112638},
  url = {https://www.sciencedirect.com/science/article/pii/S0164121225003073},
  author = {Hüseyin Ünlü and Samet Tenekeci and Dhia Eddine Kennouche and Onur Demirörs}
}

@inproceedings{Vaithilingam2022,
  author = {Vaithilingam, P. and Zhang, T. and Glassman, E. L.},
  title = {Expectation vs. experience: Evaluating the usability of code generation tools powered by large language models},
  booktitle = {Chi conference on human factors in computing systems extended abstracts},
  pages = {1--7},
  year = {2022}
}

@article{Wagner2019,
  author = {Wagner, S. and Fernández, D. M. and Felderer, M. and Vetrò, A. and Kalinowski, M. and Wieringa, R. and others},
  title = {Status quo in requirements engineering: A theory and a global family of surveys},
  journal = {ACM Transactions on Software Engineering and Methodology (TOSEM)},
  volume = {28},
  number = {2},
  pages = {1--48},
  year = {2019}
}

@incollection{Wagner2020,
  author = {Wagner, S. and Mendez, D. and Felderer, M. and Graziotin, D. and Kalinowski, M.},
  title = {Challenges in survey research},
  booktitle = {Contemporary Empirical Methods in Software Engineering},
  pages = {91--125},
  year = {2020}
}

@inproceedings{Wang2024,
  author = {Wang, R. and Cheng, R. and Ford, D. and Zimmermann, T.},
  title = {Investigating and designing for trust in AI-powered code generation tools},
  booktitle = {Proceedings of the 2024 ACM Conference on Fairness, Accountability, and Transparency},
  pages = {1475--1493},
  year = {2024}
}

@inproceedings{Zhang2023,
  author = {Zhang, K. and Wang, D. and Xia, J. and Wang, W. Y. and Li, L.},
  title = {Algo: Synthesizing algorithmic programs with generated oracle verifiers},
  booktitle = {Advances in Neural Information Processing Systems},
  volume = {36},
  pages = {54769--54784},
  year = {2023}
}

@inproceedings{Ziegler2022,
  author = {Ziegler, A. and Kalliamvakou, E. and Li, X. A. and Rice, A. and Rifkin, D. and Simister, S. and others},
  title = {Productivity assessment of neural code completion},
  booktitle = {Proceedings of the 6th ACM SIGPLAN International Symposium on Machine Programming},
  pages = {21--29},
  year = {2022}
}

@inproceedings{sikand2024much,
  title={How much space do metrics have in genai assisted software development?},
  author={Sikand, Samarth and Phokela, Kanchanjot Kaur and Sharma, Vibhu Saujanya and Singi, Kapil and Kaulgud, Vikrant and Tung, Teresa and Sharma, Pragya and Burden, Adam P},
  booktitle={Proceedings of the 17th Innovations in Software Engineering Conference},
  pages={1--5},
  year={2024}
}

@article{mastropaolo2024rise,
  title={The rise and fall (?) of software engineering},
  author={Mastropaolo, Antonio and Escobar-Vel{\'a}squez, Camilo and Linares-V{\'a}squez, Mario},
  journal={arXiv preprint arXiv:2406.10141},
  year={2024}
}

@article{schneider2024governance,
  title={Governance of generative artificial intelligence for companies},
  author={Schneider, Johannes and Kuss, Pauline and Abraham, Rene and Meske, Christian},
  journal={arXiv preprint arXiv:2403.08802},
  year={2024}
}

@inproceedings{kafoe2025making,
  title={" Making Our Life Less Monotonous" or" Just Tick Things Off": An Exploratory Multi-Method Study of Toil},
  author={Kafoe, Tom and Venegas, Lina Ochoa and Siravuru, Sharath and Serebrenik, Alexander},
  booktitle={48th International Conference on Software Engineering},
  year={2025}
}

@article{ciriello2023yea,
  title={The yea-paradox: cognitive bias in technology acceptance surveys},
  author={Ciriello, Raffaele F and Loss, Sebastian},
  journal={Information and Software Technology},
  volume={161},
  pages={107253},
  year={2023},
  publisher={Elsevier}
}

@article{schaupp2010file,
  title={E-file adoption: A study of US taxpayers’ intentions},
  author={Schaupp, Ludwig Christian and Carter, Lemuria and McBride, Megan E},
  journal={Computers in Human Behavior},
  volume={26},
  number={4},
  pages={636--644},
  year={2010},
  publisher={Elsevier}
}

@misc{yu2026cognitiveoffloadingspeedupillusion,
      title={Cognitive offloading and the speedup illusion in human-AI interaction}, 
      author={Sunny Yu and Myra Cheng and Ahmad Jabbar and Ilia Sucholutsky and Katherine M. Collins and Dan Jurafsky and Robert D. Hawkins},
      year={2026},
      eprint={2605.23177},
      archivePrefix={arXiv},
      primaryClass={cs.CY},
      url={https://arxiv.org/abs/2605.23177}, 
}

@techreport{Bain2025,
  author = {{Bain \& Company}},
  title = {Technology Report 2025},
  url = {https://www.bain.com/globalassets/noindex/2025/bain_report_technology_report_2025.pdf},
  note = {Accessed: December 2025},
  year = {2025}
}

@article{Becker2025,
  title={Measuring the impact of early-2025 AI on experienced open-source developer productivity},
  author={Becker, Joel and Rush, Nate and Barnes, Elizabeth and Rein, David},
  journal={arXiv preprint arXiv:2507.09089},
  year={2025}
}

@techreport{Capgemini2024,
  author = {{Capgemini Research Institute}},
  title = {Turbocharging software with Gen AI},
  url = {https://www.capgemini.com/wp-content/uploads/2024/07/CRI_Turochanging-Software_Final.pdf},
  note = {Accessed: December 2025},
  year = {2024}
}

@techreport{Challapally2025,
  author = {Challapally, A. and Pease, C. and Raskar, R. and Chari, P.},
  title = {The GenAI Divide, State of AI in Business 2025},
  institution = {MIT},
  month = {July},
  url = {https://mlq.ai/media/quarterly_decks/v0.1_State_of_AI_in_Business_2025_Report.pdf},
  note = {Accessed: December 2025},
  year = {2025}
}

@online{Chandonnet2025,
  author = {Chandonnet, H.},
  title = {When Pinterest needs new AI tools, employees can have a part in creating them},
  journal = {Business Insider},
  url = {https://www.businessinsider.com/pinterest-hackathons-makeathon-employees-lead-generative-ai-tools-innovation-2025-5},
  note = {Accessed: December 2025},
  year = {2025}
}

@article{DeCampos2024,
  author = {de Campos, A. and Melegati, J. and Nascimento, N. and Chanin, R. and Sales, A. and Wiese, I.},
  title = {Some things never change: how far generative AI can really change software engineering practice},
  journal = {arXiv preprint arXiv:2406.09725},
  year = {2024}
}

@online{DefineX2025,
  author = {{DefineX}},
  title = {Beyond Code: GenAI’s Strategic Impact on the Entire Software Development Lifecycle},
  url = {https://www.teamdefinex.com/insights/beyond-code-genais-strategic-impact-on-the-entire-software-development-lifecycle/},
  note = {Accessed: December 2025},
  year = {2025}
}

@techreport{EuropeanCommission2024,
  author = {{European Commission}},
  title = {Artificial Intelligence and the future of work},
  url = {https://europa.eu/eurobarometer/surveys/detail/3222},
  note = {Accessed: December 2025},
  year = {2024}
}

@online{Gartner2024,
  author = {{Gartner}},
  title = {Gartner Says Generative AI will Require 80\% of Engineering Workforce to Upskill Through 2027},
  month = {October},
  url = {https://www.gartner.com/en/newsroom/press-releases/2024-10-03-gartner-says-generative-ai-will-require-80-percent-of-engineering-workforce-to-upskill-through-2027},
  note = {Accessed: December 2025},
  year = {2024}
}

@techreport{Google2025a,
  author = {{Google}},
  title = {DORA Report - Impact of generative AI in software development},
  url = {https://dora.dev/ai/gen-ai-report/},
  note = {Accessed: December 2025},
  year = {2025}
}

@techreport{Google2025b,
  author = {{Google}},
  title = {DORA Report - State of AI-assisted Software Development},
  url = {https://dora.dev/research/2025/dora-report/},
  note = {Accessed: December 2025},
  year = {2025}
}

@article{Guo2025,
  author = {Guo, L. and Ye, S. and Sun, Z. and Chen, X. and Zhang, Y. and Wang, B. and others},
  title = {LLMs are Bug Replicators: An Empirical Study on LLMs’ Capability in Completing Bug-prone Code},
  journal = {arXiv preprint arXiv:2503.11082},
  year = {2025}
}

@techreport{Handa2025,
  author = {Handa, K. and Tamkin, A. and McCain, M. and Huang, S. and Durmus, E. and Heck, S. and others},
  title = {Which Economic Tasks are Performed with AI? Evidence from Millions of Claude Conversations},
  institution = {Anthropic},
  url = {https://assets.anthropic.com/m/2e23255f1e84ca97/original/Economic_Tasks_Al_Paper.pdf},
  note = {Accessed: December 2025},
  year = {2025}
}

@article{Haque2024,
  author = {Haque, E. A. and Brown, C. and LaToza, T. D. and Johnson, B.},
  title = {Information seeking using AI assistants},
  journal = {arXiv preprint arXiv:2408.04032},
  year = {2024}
}

@article{Jiang2024,
  author = {Jiang, J. and Wang, F. and Shen, J. and Kim, S. and Kim, S.},
  title = {A survey on large language models for code generation},
  journal = {arXiv preprint arXiv:2406.00515},
  year = {2024}
}

@online{Kästner2024,
  author = {Kästner, C.},
  title = {Software Engineering for AI/ML - An Annotated Bibliography},
  url = {https://github.com/ckaestne/seaibib},
  note = {Accessed: December 2025},
  year = {2024}
}

@techreport{McKinsey2023,
  author = {{McKinsey \& Company}},
  title = {The economic potential of generative AI},
  month = {June},
  url = {https://www.mckinsey.com/capabilities/tech-and-ai/our-insights/the-economic-potential-of-generative-ai-the-next-productivity-frontier},
  note = {Accessed: December 2025},
  year = {2023}
}

@online{Muratovic2024,
  author = {Muratovic, F. and Gill, J. and Kearns-Manolatos and Alibage, A.},
  title = {How can organizations engineer quality software in the age of generative AI?},
  institution = {Deloitte},
  month = {October},
  url = {https://www.deloitte.com/us/en/insights/industry/technology/how-can-organizations-develop-quality-software-in-age-of-gen-ai.html},
  note = {Accessed: December 2025},
  year = {2024}
}

@techreport{OpenAI2025,
  author = {{OpenAI}},
  title = {ChatGPT usage and adoption patterns at work},
  url = {https://cdn.openai.com/pdf/3c7f7e1b-36c4-446b-916c-11183e4266b7/chatgpt-usage-and-adoption-patterns-at-work.pdf},
  note = {Accessed: December 2025},
  year = {2025}
}

@techreport{Ozkaya2025,
  author = {Ozkaya, I. and Carleton, A. and Butkovic, M. and Echeverria, S. and Edman, R. and Haller, J. and others},
  title = {A Preliminary Report on a Model for Maturing AI Adoption: From Hype to Achieving Repeatable, Predictable Outcomes},
  institution = {Carnegie Mellon University, Software Engineering Institute},
  month = {December},
  url = {https://www.sei.cmu.edu/documents/6413/AI-Maturity-Model-Overview.pdf},
  note = {Accessed: December 2025},
  year = {2025}
}

@article{Peng2023,
  title={The impact of ai on developer productivity: Evidence from github copilot},
  author={Peng, Sida and Kalliamvakou, Eirini and Cihon, Peter and Demirer, Mert},
  journal={arXiv preprint arXiv:2302.06590},
  year={2023}
}

@article{Song2024,
  author = {Song, F. and Agarwal, A. and Wen, W.},
  title = {The impact of generative AI on collaborative open-source software development: Evidence from GitHub Copilot},
  journal = {arXiv preprint arXiv:2410.02091},
  year = {2024}
}

@online{StackOverflow2025,
  author = {{StackOverflow}},
  title = {Developer Survey 2025},
  url = {https://survey.stackoverflow.co/2025/ai/},
  note = {Accessed: December 2025},
  year = {2025}
}

@techreport{StanfordHAI2025,
  author = {{Stanford HAI}},
  title = {Artificial Intelligence Index Report 2025},
  url = {https://hai.stanford.edu/assets/files/hai_ai_index_report_2025.pdf},
  note = {Accessed: December 2025},
  year = {2025}
}

@online{Laura2025,
  author = {Tacho, Laura},
  title = {AI-assisted engineering: Q4 impact report},
  url = {https://getdx.com/blog/ai-assisted-engineering-q4-impact-report-2025/},
  note = {Accessed: December 2025},
  year = {2025}
}

@techreport{WEF2025,
  author = {{World Economic Forum}},
  title = {The Future of Jobs Report 2025},
  url = {https://www.weforum.org/publications/the-future-of-jobs-report-2025/},
  note = {Accessed: December 2025},
  year = {2025}
}

@article{coughlin1990recall,
  title={Recall bias in epidemiologic studies},
  author={Coughlin, Steven S},
  journal={Journal of clinical epidemiology},
  volume={43},
  number={1},
  pages={87--91},
  year={1990},
  publisher={Elsevier}
}

@article{krumpal2013determinants,
  title={Determinants of social desirability bias in sensitive surveys: a literature review},
  author={Krumpal, Ivar},
  journal={Quality \& quantity},
  volume={47},
  number={4},
  pages={2025--2047},
  year={2013},
  publisher={Springer}
}

@article{mabe1982validity,
  title={Validity of self-evaluation of ability: A review and meta-analysis.},
  author={Mabe, Paul A and West, Stephen G},
  journal={Journal of applied Psychology},
  volume={67},
  number={3},
  pages={280},
  year={1982},
  publisher={American Psychological Association}
}

@inproceedings{tomaz2026impacts,
  title={Impacts of Generative AI on Agile Teams' Productivity: A Multi-Case Longitudinal Study},
  author={Tomaz, Rafael and Guenes, Paloma and Ara{\'u}jo, Allysson Allex and Baldassarre, Maria Teresa and Kalinowski, Marcos},
  booktitle={Proceedings of the 2026 IEEE/ACM Third International Conference on AI Foundation Models and Software Engineering},
  pages={128--138},
  year={2026}
}

@article{tabarsi2027llms,
  title={LLMs’ reshaping of people, processes, products, and society in software development: a qualitative exploration with early adopters},
  author={Tabarsi, Benyamin and Reichert, Heidi and Gilson, Sam and Limke, Ally and Kuttal, Sandeep and Barnes, Tiffany},
  journal={Empirical Software Engineering},
  volume={32},
  number={1},
  pages={21},
  year={2027},
  publisher={Springer}
}

@inproceedings{he2026speed,
  title={Speed at the cost of quality: How Cursor AI increases short-term velocity and long-term complexity in open-source projects},
  author={He, Hao and Miller, Courtney and Agarwal, Shyam and K{\"a}stner, Christian and Vasilescu, Bogdan},
  booktitle={Proceedings of the 23rd international conference on mining software repositories},
  pages={181--193},
  year={2026}
}

@inproceedings{huang2026more,
  title={More Code, Less Reuse: Investigation on Code Quality and Reviewer Sentiment towards AI-generated Pull Requests},
  author={Huang, Haoming and Jaisri, Pongchai and Shimizu, Shota and Chen, Lingfeng and Nakashima, Sota and Rodr{\'\i}guez-P{\'e}rez, Gema},
  booktitle={Proceedings of the 23rd International Conference on Mining Software Repositories},
  pages={1024--1028},
  year={2026}
}

@article{cui2026effects,
  title={The effects of generative AI on high-skilled work: Evidence from three field experiments with software developers},
  author={Cui, Kevin Zheyuan and Demirer, Mert and Jaffe, Sonia and Musolff, Leon and Peng, Sida and Salz, Tobias},
  journal={Management Science},
  year={2026},
  publisher={INFORMS}
}

@techreport{demirer2026writing,
  title={Writing code vs. shipping code: Productivity effects across generations of AI coding tools},
  author={Demirer, Mert and Musolff, Leon and Yang, Liyuan},
  year={2026},
  institution={National Bureau of Economic Research}
}

@inproceedings{ulloa2026product,
  title={Product Manager Practices for Delegating Work to Generative AI:“Accountability must not be delegated to non-human actors”},
  author={Ulloa, Mara and Butler, Jenna L and Haniyur, Sankeerti and Miller, Courtney and Amos, Barrett and Sarkar, Advait and Storey, Margaret-Anne},
  booktitle={Proceedings of the IEEE/ACM 48th International Conference on Software Engineering: Software Engineering in Practice},
  pages={10--21},
  year={2026}
}

@article{diebold2025backlogs,
  title={From backlogs to bots: Generative AI's impact on agile role evolution},
  author={Diebold, Philipp},
  journal={Journal of Software: Evolution and Process},
  volume={37},
  number={1},
  pages={e2740},
  year={2025},
  publisher={Wiley Online Library}
}

\appendix
\section{Selected Participant Responses and Assigned Categories for SE Tasks, Benefits, and Challenges}
\label{sec:appendix}

\setcounter{table}{0}
\renewcommand{\thetable}{A.\arabic{table}}
\scriptsize
\begin{longtable}{>{\raggedright\arraybackslash}p{0.28\linewidth} >{\raggedright\arraybackslash}p{0.67\linewidth}}
\caption{Mapping of SE knowledge areas to selected participant responses} \label{tab:se_knowledge_areas} \\
\toprule
\textbf{SE Knowledge Area} & \textbf{Selected Responses} \\
\midrule
\endfirsthead

\multicolumn{2}{c}{{\bfseries \tablename\ \thetable{} -- Continued from previous page}} \\
\toprule
\textbf{SE Knowledge Area} & \textbf{Selected Responses} \\
\midrule
\endhead

\midrule
\multicolumn{2}{r}{{Continued on next page}} \\
\bottomrule
\endfoot

\bottomrule
\endlastfoot

Acquisition \& Supply & Tender \\
Architecture Definition & Architecture, architecture design, architecture planning \\
Business or Mission Analysis & Market analysis, gather market information \\
Design Definition & Design, solution design, database design, database modeling, API design, interface design \\
Implementation & Coding, development, programming, software development, application development, code generation, method generation, script generation, SQL generation, code completion, unit testing, prototyping, product development \\
Information Management & Documentation, technical documentation, documentation generation, writing documents, drafting documents, diagram creation, presentation \\
Integration & System integration \\
Maintenance & Understanding code, code explanation, code summarization, code modification, code optimization, database optimization, debugging, bug analysis, bug fixing, refactoring, code cleanup \\
Measurement & Data analysis \\
Operation & Troubleshooting, incident management \\
Personal Assistance & Brainstorming, inspiration, research, search, proof reading, summarization, question answering, learning, writing, organizing thoughts, communication, training, problem solving \\
Project Assessment \& Control & Project management refinement \\
Project Planning & Planning, task planning, estimation, generate project timeline \\
Quality Assurance & QA, quality assurance, code fix \\
Requirements Definition & Requirements engineering, requirements gathering, requirements elicitation, requirements analysis, requirements definition, requirements validation, mockup, user story writing, user story refinement, spike development \\
Transition & DevOps, CI/CD \\
Verification \& Validation & Testing, test generation, test case generation, test case writing, test data generation, code review, PR review, validation \\
\end{longtable}

\scriptsize
\begin{longtable}{>{\raggedright\arraybackslash}p{0.28\linewidth} >{\raggedright\arraybackslash}p{0.67\linewidth}}
\caption{Mapping of benefit categories to selected participant responses} \label{tab:perceived_benefits} \\
\toprule
\textbf{Benefit Category} & \textbf{Selected Responses} \\
\midrule
\endfirsthead

\multicolumn{2}{c}{{\bfseries \tablename\ \thetable{} -- Continued from previous page}} \\
\toprule
\textbf{Benefit Category} & \textbf{Selected Responses} \\
\midrule
\endhead

\midrule
\multicolumn{2}{r}{{Continued on next page}} \\
\bottomrule
\endfoot

\bottomrule
\endlastfoot

Help to eliminate repetitive tasks & Eliminate/reduce/automate repetitive/routine tasks \\
Improve communication & Better/improved communication, coordination \\
Improve quality & Quality, fewer bugs, reducing errors, good documentation, grammar correction, accuracy, improved code quality, improve results, quality of output \\
Increase productivity & Increased/enhanced/high productivity, efficient/efficiency, more productive, improved/increased work efficiency \\
Reduce cycle-time & Time saving, time reduction, faster development/coding/debugging/prototyping, speed, fast \\
Support conducting a SE task & Code creation/completion/drafting, create basic project tasks, generate knowledge base, write script, write unit test, generate boilerplate code \\
Support creativity & Find new ideas, brainstorming, idea generation, explore alternative solutions, idea validation, innovation \\
Support knowledge search & Knowledge search, research, faster search \\
Support learning & Learning, find best practices, learning new concepts/technologies, understanding existing code/codebase \\
Support problem solving & Problem solving, error explanation, understand complex code, bug identification and resolution, getting suggestions to fix errors, suggestions for root cause \\
\end{longtable}

\scriptsize
\begin{longtable}{>{\raggedright\arraybackslash}p{0.28\linewidth} >{\raggedright\arraybackslash}p{0.67\linewidth}}
\caption{Mapping of challenge categories to selected participant responses} \label{tab:genai_challenges} \\
\toprule
\textbf{Challenge Category} & \textbf{Selected Responses} \\
\midrule
\endfirsthead

\multicolumn{2}{c}{{\bfseries \tablename\ \thetable{} -- Continued from previous page}} \\
\toprule
\textbf{Challenge Category} & \textbf{Selected Responses} \\
\midrule
\endhead

\midrule
\multicolumn{2}{r}{{Continued on next page}} \\
\bottomrule
\endfoot

\bottomrule
\endlastfoot

Access limitations to GenAI tools & I believe that access to ChatGPT, for instance, would enable me to further explore the potential applications of large language models \\
Adoption resistance & Resistance from some team members \\
Adoption to evolving GenAI tools & Adapting as GenAI tools are evolving; to keep with all the new tools \\
Copyright and IP concerns & IP concerns; concerns about intellectual property \\
Cost & Price \\
Forgetting & Long chats lack the needed continuity; next time it forgets; limited context window; occasionally loses track of earlier decisions or constraints \\
Inaccurate output/Hallucination & Hallucination; accuracy; inaccurate; wrong answer/suggestion; obvious mistakes; fake resource/reference; useless code \\
Integration with existing environment & Lack of integration between platforms; potential integration issues within existing workflows; integration with existing corporate systems can be challenging; integration into existing workflow not seamless \\
Interaction with GenAI tool & Doesn't understand me very well; sometimes the model doesn't understand the prompt; it is difficult to create effective prompts; ignoring my instructions; making myself clear; knowing how to write a prompt; making the tool understand what I need exactly; adjust input commands \\
Lack of context understanding & Lack of context; lack of detailed knowledge; limited context awareness; lack of domain-specific context; limited understanding of project-specific requirements \\
Legal/audit compliance concerns & Legal challenges; compliance and audit concerns \\
Output quality issues & Sometimes doesn't tell more than what's already obvious; clarity of requirements; code style incoherent; responses can be too high-level or abstract; suggestions that are inefficient or not optimized \\
Output reliability and consistency & Results are either almost perfect or complete useless garbage; inconsistent performance; reliability; less reliable answers \\
Output validation and review & Difficulty in evaluating the correctness of generated code; I have to manually find the issue to fix it; a manual review will be needed to incorporate in code; review and crosscheck; analyze and validate all results; need for extra time to review and validate; need for careful human validation; need for human review and refinement \\
Platform and infrastructure readiness & Not applicable on some platforms; platform challenges \\
Privacy and ethical concerns & Data privacy; concerns about data privacy \\
Risk of overreliance & Overreliance; loosing the critical thinking on the output of the models \\
Security concerns & Security concerns; data security; security challenges \\
Tool selection & To distinguish when to use the tools and when not to \\
Unwanted suggestions & Completing code when not intending to; invasive autocomplete \\
\end{longtable}

\end{document}